\documentclass{aa}  
\usepackage{graphicx}
\usepackage{txfonts}
\usepackage{caption}
\usepackage{multicol}       
\usepackage{multirow}
\usepackage[utf8]{inputenc}
\usepackage{booktabs}
\usepackage{natbib}
\bibpunct{(}{)}{;}{a}{}{,} 

\usepackage[colorlinks=true, urlcolor=blue, colorlinks=true, citecolor=blue, linkcolor=blue]{hyperref}

\begin{document} 

   \title{The Tremaine-Weinberg method at high redshifts}
   
   \author{Mahmood Roshan
          \inst{1}\thanks{\email{mroshan@um.ac.ir}} \and Asiyeh Habibi \inst{1} \and J. Alfonso L. Aguerri \inst{2,}\inst{3} \and Virginia Cuomo \inst{4} \and Connor Bottrell \inst{5} \and Luca Costantin \inst{6} \and Enrico Maria Corsini \inst{7,}\inst{8}  \and Taehyun Kim \inst{9} \and Yun Hee Lee \inst{9,10} \and Jairo Mendez-Abreu \inst{3,}\inst{2} \and Matthew Frosst \inst{5} \and Adriana de Lorenzo-Cáceres\inst{3,}\inst{2} \and Lorenzo Morelli \inst{11} \and Alessandro Pizzella\inst{7,}\inst{8}
          }

   \institute{Department of Physics, Faculty of Science, Ferdowsi University of Mashhad, P.O. Box 1436, Mashhad, Iran 
              \and Instituto de Astrofísica de Canarias, Calle Vía Láctea s/n, E-38205 La Laguna, Tenerife, Spain \and Departamento de Astrofísica de la Universidad de La Laguna, Av Astrofisíco Francisco Sánchez s/n, 38205 La Laguna, Tenerife, Spain \and Departamento de Astronomía, Universidad de La Serena, Av. Raúl Bitrán 1305, La Serena, Chile \and International Centre for Radio Astronomy Research, University of Western Australia, 35 Stirling Hwy, Crawley, WA 6009, Australia \and Centro de Astrobiolog{\'{\i}}a, CSIC-INTA, Ctra. de Ajalvir km 4, Torrej{\'o}n de Ardoz, E-28850,
Madrid, Spain \and Dipartimento di Fisica e Astronomia “G. Galilei”, Università di
Padova, vicolo dell’Osservatorio 3, I-35122 Padova, Italy \and INAF – Osservatorio Astronomico di Padova, Vicolo dell’Osservatorio 5, I-35122 Padova, Italy \and Department of Astronomy and Atmospheric Sciences, Kyungpook National University, Daegu 41566, Republic of Korea
\and Korea Astronomy and Space Science Institute (KASI), 776 Daedeokdae-ro, Yuseong-gu, Daejeon 34055, Republic of Korea \and Instituto de Astronomíay Ciencias Planetarias, Universidad de Atacama, Avenida Copayapu 485, 1530000 Copiapó, Atacama, Chile
             }

   \date{\today}

  \abstract
{This paper examines the reliability of the Tremaine-Weinberg (TW) method in measuring the pattern speed of barred galaxies at high redshifts. Measuring pattern speeds at high redshift may help to shed light on the time evolution of interactions between galactic bars and dark matter halos. The TW method has been extensively employed for nearby galaxies, and its accuracy in determining bar pattern speeds has been validated through numerical simulations. For nearby galaxies, the method yields acceptable results when the inclination angle of the galaxy and the position angle of the bar fall within appropriate ranges. However, the application of the TW method to high-redshift galaxies remains unexplored in both observations and simulations. For this study we generated mock observations of barred galaxies from the TNG50 cosmological simulation. These simulated observations were tailored to mimic the integral
field unit (IFU) spectroscopy data that the Near-Infrared Spectrograph (NIRSpec) on the James Webb Space Telescope (JWST) would capture at a redshift of $z\simeq 1.2$. By applying the TW method to these mock observations and comparing the results with the known pattern speeds, we demonstrate that the TW method performs adequately for barred galaxies whose bars are sufficiently long to be detected by JWST at high redshifts. This work opens a new avenue for applying the TW method to investigate the properties of high-redshift barred galaxies.} 

   \keywords{galaxies: spiral -- galaxies: evolution -- galaxies: kinematics and dynamics -- galaxies: structure}

   \maketitle

\section{Introduction}
\label{intro}

Barred galaxies provide a rich framework for understanding the structural and dynamical evolution of disk galaxies. These prominent structures, formed by the collective motion of stars within galactic disks, play a critical role in redistributing angular momentum, channeling gas inflows, and influencing the growth of bulges and supermassive black holes \citep{2009ASSP....8..159G,math2025}. The bar pattern speed ($\Omega_{\rm p}$), which quantifies the angular velocity of the bar, is a fundamental parameter in understanding these processes. It governs the interactions between bars and dark matter halos and the formation of resonances \citep{at1992, Athanassoula2013, 2014RvMP...86....1S}. 

Despite significant progress in studying bar pattern speeds in nearby galaxies, their evolution at higher redshifts remains poorly understood.  The dimensionless parameter $\mathcal{R}$ is conveniently defined as the ratio of the corotation radius ($R_{\rm cr}$) to the bar radius ($R_{\rm bar}$), i.e., $ \mathcal{R} = R_{\rm cr}/R_{\rm bar}$. The corotation radius is related to the pattern speed $\Omega_{\rm p}$ by the equation $\Omega(R_{\rm cr})=\Omega_{\rm p}$, where $\Omega$ represents the angular velocity derived from the circular velocity of the stars, given by $v_{\rm cir}=R \Omega(R)$. Cosmological simulations, such as EAGLE \citep{eagle} and IllustrisTNG  \citep{tng1,tng2,tng3,tng4,tng5,TNG502,TNG501,TNG503}, predict that most bars are slow ($\mathcal{R} > 1.4$), while observations of local galaxies suggest that a considerable fraction of bars are fast ($\mathcal{R} < 1.4$) \citep{Aguerri2015, algorry2017, Cuomo2020, 2021MNRAS.508..926R}. It is important to mention that \cite{Geron2023} measured the pattern speed of 225 barred galaxies.\footnote{While \cite{Geron2023} report pattern speed measurements for 225 barred galaxies, it is likely that only a subset of these have high-accuracy determinations of $\Omega$. Notably, among the 50 randomly selected galaxies presented in their Table 3, more than 40\% of the measured pattern speeds exhibit relative errors exceeding 30\%.} This study benefits from a larger sample of galaxies compared to previous studies. The results show that about 62\% of the bars in their sample are slow. However, the mean value of $\mathcal{R}$ is approximately 1.7, which remains significantly different from the TNG50 value of about 3.1. This discrepancy, referred to as the fast bar tension, raises critical questions about angular momentum exchange between bars and their surrounding halos, as well as the influence of baryonic processes on bar dynamics. If the pattern speed of a bar slows down due to dynamical friction between the bar and dark matter halo, this discrepancy raises questions about the nature and distribution of dark matter within galaxies \citep{debattista2000, 2021A&A...650L..16F}. This also fuels motivations for pattern speed measurements at high redshifts.

Various indirect methods have been devised to estimate $\Omega_{\rm p}$, each offering distinct advantages and constraints. Resonance-based analyses identify rings associated with specific resonances, such as the inner Lindblad resonance (ILR), the corotation resonance (CR), or the outer Lindblad resonance (OLR) to infer pattern speeds \citep [e.g.,][]{Buta1986,Buta1995a,Mu2004}. This resonance-based method has been used to measure $\Omega_{\rm p}$ for galaxies at intermediate redshifts \citep{2012A&A...540A.103P}. By developing a mass model of the galaxy, one can determine its circular velocity and epicyclic frequency. This approach enables the identification of resonance locations, especially the $R_{\rm cr}$, which in turn allows  the measurement of the pattern speed. While mathematically straightforward, this method relies heavily on the accurate   identification of ring structures and galaxy mass modeling, which can be challenging in galaxies with complex or ambiguous morphologies \citep[e.g.,][]{kormendy1979, buta1995, jeong2007}. 

Interestingly, as another indirect approach, the location near dark gaps (i.e., regions within a barred galaxy where the difference in surface brightness between the major and minor axes of the bar drops to zero after reaching its maximum)  has been used to determine the corotation radius \citep{Krishnarao2022}. Residuals in observed velocity maps can also help determine the corotation radius. This method relies on the phase change of the noncircular radial velocity across corotation \citep{Sempere1995,Font2011, Font2014, Font2017,Beckman2018}.

When a bar density wave is present in the galactic disk, a phase shift occurs between the density wave and potential wave. The radial distribution of this potential-density phase shift is characterized by a change in sign at the corotation radius \citep{1998ApJ...499...93Z}. This method, which is independent of kinematic data, is used to find the location of the corotation radius, and consequently the pattern speed \citep{zhang2007}. 

\cite{at1992} via hydrodynamical simulations proposed a method for determining the corotation radius by analyzing the offset and shape of dust lanes, which are indicators of the ILR. This approach requires high-resolution imaging coupled with  an accurate interpretation of dust lane patterns. 

Another indirect approach to measure the pattern speed involves N-body or hydrodynamical modeling of barred galaxies. This method entails constructing a comprehensive model of  gravitational potential of the galaxy. The baryonic component is modeled using photometric data, while the dark matter halo potential can be fitted using rotation curve observations. By simulating the response of star and gas particles to a rotating rigid bar within this model, one can infer the pattern speed \citep{england1990,lindblad1996, Aguerri2001, weiner2001,Rautiainen2008,lin2013}. 

The Tremaine-Weinberg (TW) method stands out as a direct and robust alternative. The TW method offers plausible estimates of $\Omega_{\rm p}$ by integrating photometric and kinematic data. However, its accuracy is contingent upon several well-defined conditions: the galactic disk must be thin, the inclination angle $i$ of the disk should fall within an intermediate range $15^{\circ}\lesssim i\lesssim 70^{\circ}$, the position angle of the bar relative to the major axis of the galaxy $\Delta PA$ must be within an appropriate interval $10^{\circ}\lesssim \Delta PA\lesssim 75^{\circ}$, and there should not be significant star formation in the disk \citep{Tremaine1984,corsini2011,Zou2019}. The TW method has been extensively applied to observations of nearby galaxies \citep{Deb2002,Aguerri2003,Aguerri2015,Deb2004,Corsini2005,Gerssen2007,corsini2011,Cuomo2019b,Cuomo2019a,Cuomo2020,Cuomo2021,Cuomo2022,Guo2019,Garma2020,Buttitta2022,Buttitta2024, Garma2022,Geron2023}. The introduction of integral field spectroscopy (IFS) has significantly expanded the sample of galaxies to which the TW method can be applied. This technological advancement has enabled us to employ the TW method across a broader range of galactic types, encompassing both early-type and late-type disk galaxies. Moreover, the method has found application in cosmological simulations, including projects such as IllustrisTNG \citep{2021MNRAS.508..926R,AHabibi}. However, to date this method has  only been applied to low-redshift observations ($z < 0.1$), leaving its application to high-redshift galaxies unexplored. Poor spatial resolution in distant galaxies may complicate its implementation, though no systematic studies have yet addressed this challenge. Even intermediate redshifts ($z \sim 0.5$) remain untested, creating significant gaps in our understanding of bar dynamics at earlier cosmic epochs. It is worth noting that barred galaxies do exist at high redshifts, but this raises questions about how they form at such early times \citep{bar_2023,2023Natur.623..499C,mobasher2023,2023A&A...678A..54M,bar2024_1,bar2024_2,2025MNRAS.537.1163A,2025arXiv250321738E,2025arXiv250203532H,luc1,luc2}.

The aim of this study is to assess the efficacy of the TW method for barred galaxies at a redshift of approximately 1.2. We utilized mock observations derived from the TNG50 cosmological simulation to create mock data that emulate the IFS capabilities of the Near-Infrared Spectrograph (NIRSpec) aboard the JWST. By comparing TW-derived pattern speeds with baseline values obtained using the Dehnen-Semczuk-Schönrich (DSS) method \citep{newps}, we assessed the accuracy and limitations of the TW method under realistic observational conditions. The DSS method is an unbiased method to measure the bar pattern speed from single simulation snapshots. Our findings will not only bridge the gap between simulations and high-redshift observations, but will also lay the groundwork for future observational campaigns aimed at exploring bar dynamics across cosmic time.

Given the crucial role of hydrodynamical simulations in understanding galaxy structure and evolution, generating synthetic observations is an essential step for comparing simulated results with real observational data. It is worth noting that there are at least two distinct methods for generating mock observations from controlled or cosmological hydrodynamical simulations. The most realistic approach involves using radiative transfer to simulate the effects of the interstellar medium, thereby producing synthetic spatially resolved spectra for hydrodynamical simulations. Subsequently, other kinematic properties can be extracted through spectral analysis. This approach has been widely implemented for creating mock observations that simulate various instrumental setups, as demonstrated by several studies \citep{Camps2016,nanni2022,2023MNRAS.524..907B,harborne2023,2023ApJ...946...71C,Baes2024,2025arXiv250618997I}. 

The second approach, which is adopted in this paper, offers a computationally efficient alternative that circumvents the need for radiative transfer and spectral fitting by directly constructing line-of-sight velocity distribution (LOSVD) cubes, as detailed in Section \ref{losvd}. Instead of generating frequency cubes, velocity cubes are constructed. This approach significantly reduces the computational resources required while still providing valuable kinematic insights \citep{connor2022}.

The paper is structured as follows. Section \ref{TW-method} provides a concise overview of the mathematical foundations underlying the TW method. Section \ref{Data} introduces the simulated galaxies from TNG50, detailing our methodology for selecting barred galaxies and measuring their bar length and strength, and describing the process of constructing LOSVD cubes. In Section \ref{comp} we utilize both the TW and DSS methods on high-resolution TNG50 galaxies to demonstrate the accuracy of the TW approach. Section \ref{Synthetic} complements Section \ref{Data} by detailing the process of creating the final mock galaxies from TNG50 data. Specifically, this section focuses on how the instrumental setup is incorporated into the mock data. Section \ref{highz} applies the TW method to our mock observations at redshift $z\simeq 1.2$. Section \ref{extra} provides an additional test, treating the mock observations as real galaxies and measuring their bar position angle and disk inclination using ellipse fitting techniques. Section \ref{discussion} presents our results and conclusions. Appendix \ref{test} compares the TW and DSS methods using an isolated disk simulation, establishing the DSS method as a baseline. Appendix \ref{JWST_TW} explores how far we can apply the TW methods to obtain $\Omega_{\mathrm{p}}$ and $R_{\rm CR}$ with the JWST. Throughout this paper, we adopt cosmological parameters consistent with those used in the IllustrisTNG simulation: $H_{0} = 67.74 \, \rm{km\,s^{-1}\, Mpc^{-1}}$, $\Omega_{\mathrm{b},0} = 0.0486$, $\Omega_{\mathrm{m},0} = 0.3089$, $\Omega_{\mathrm{\Lambda},0} = 0.6911$.

\section{The Tremaine-Weinberg method}
\label{TW-method}

The TW method is the primary technique for measuring the bar pattern speed, $\Omega_{\rm p}$. To apply this method, we must select a tracer population of stars or gas that satisfies the continuity equation, with
i) mass surface density $\Sigma(X,Y)$, where $(X,Y)$ denotes the Cartesian coordinates on the disk, with the origin at the galaxy center, and the $X$  and $Y$ coordinates running along the galaxy major and minor axis, respectively, and ii) line-of-sight velocity $V_{\text{LOS}}(X,Y)$. The method involves multiplying the continuity equation for a thin disk by an arbitrary window function $W(Y)$ and integrating over the positions \citep{Tremaine1984}. This yields the pattern speed as
$\Omega_{\rm p} \sin i = \langle V \rangle/\langle X \rangle$
where $i$ is the galaxy inclination angle. The kinematic and photometric integrals are defined respectively as:
\begin{eqnarray}
    \langle V \rangle ~&\equiv&~ \frac{\int_{-\infty}^{\infty} \frac{dW(Y)}{dY} \int_{-\infty}^{\infty} V_{\text{LOS}}(X,Y)\, \Sigma(X,Y) \, dX\,dY}{\int_{-\infty}^{\infty} \frac{dW(Y)}{dY} \int_{-\infty}^{\infty} \Sigma(X,Y) \, dX\, dY} \, , \\
    \langle X \rangle ~&\equiv&~ \frac{\int_{-\infty}^{\infty} \frac{dW(Y)}{dY} \int_{-\infty}^{\infty} X\, \Sigma(X,Y) \, dX\,dY}{\int_{-\infty}^{\infty} \frac{dW(Y)}{dY} \int_{-\infty}^{\infty} \Sigma(X,Y) \, dX\, dY} .
\end{eqnarray}
The window function cannot depend on $X$, as this would introduce the velocity component perpendicular to the line-of-sight, which cannot be measured in observations, into the integrals. One commonly used window function is $W(Y) = \Theta(Y - Y_0)$, where $\Theta(Y - Y_0)$ represents the Heaviside step function. In this case, $dW/dY = \delta(Y - Y_0)$, where $\delta$ is the Dirac delta function, and the kinematic and photometric integrals simplify to the following form:
\begin{eqnarray}
    \langle V \rangle_{Y_0} ~&\equiv&~ \frac{ \int_{-\infty}^{\infty} V_{\text{LOS}}(X,Y_0)\, \Sigma(X,Y_0) \, dX}{ \int_{-\infty}^{\infty} \Sigma(X,Y_0) \, dX} \, , \\
    \langle X \rangle_{Y_0} ~&\equiv&~ \frac{ \int_{-\infty}^{\infty} X\, \Sigma(X,Y_0) \, dX}{ \int_{-\infty}^{\infty} \Sigma(X,Y_0) \, dX}  .
\end{eqnarray}
Here, $Y = Y_0$ represents a slit parallel to the $X$ axis, positioned at a distance $Y_0$ from the galaxy major axis. The above integrals are taken along this slit. By selecting $(2k+1)$ different values for $Y_0$, evenly distributed around the disk major axis, we obtain $(2k+1)$ points in the $\langle X\rangle - \langle V \rangle$ plane. If a bar is present in the system, these points will align along a straight line. As previously mentioned, the slope of this line yields $\Omega_{\rm p} \sin i$.

\section{Data and methods}
\label{Data}

We conduct tests and demonstrate the application of the TW method to make mock galaxies using the TNG50-1 run from the IllustrisTNG cosmological magnetohydrodynamical simulation. We selected this simulation because it offers sufficient resolution for our study and the presence and properties of stellar bars have been extensively explored within it \citep{2022MNRAS.512.5339R,2022MNRAS.515.1524Z, AHabibi,dehnen2024}. Furthermore, bars are a common feature in TNG50 galaxies, as expected. For this purpose, we select barred galaxies with stellar masses $M_* > 10^{10.0}\, M_{\odot}$ from TNG50 at redshifts $z=0, 0.5$, and $1$. This mass range is chosen to ensure sufficient resolution, or number of particles, for reliable identification of stellar bars. We do not investigate the evolution of bars with redshift in this study. For such a study, we refer to \citet{AHabibi}, \citet{dehnen2024}, and \citet{math2025}.
 Instead, we combine the barred galaxies selected from redshifts $z=0, 0.5$, and $1$ (different snapshots in TNG50), just to enlarge out galaxy sample. Then we treat them as individual, independent galaxies. We assume that these galaxies are representative of nearby galaxies. Since the simulation provides precise information about the positions and velocities of all particles, we can artificially treat them as if they were nearby galaxies. Subsequently, we make mock data cubes to replicate observations of these galaxies at their respective redshifts. This combination allows us to create a large sample of galaxies for our TW method test, leading to statistically robust conclusions about the method reliability. In the following subsections, we briefly explain our procedure for selecting and characterizing barred galaxies. More details can be found in \cite{2021MNRAS.508..926R}.

\subsection{Selecting disk galaxies}
\label{sample-selection}
To identify barred galactic disks, we first need to distinguish disk galaxies within our sample. The center of each galaxy is assumed to coincide with the center of mass of its inner regions. By inner region, we mean the volume enclosed within a sphere of radius $4\,\text{kpc}$ centered on the center of mass of the entire stellar component of the galaxy. We begin by calculating the direction of the total angular momentum vector for stellar particles within the stellar half-mass radius.\footnote{Since the center of the galaxy has been determined, we simply calculate the radius of the sphere that encloses half of the total stellar mass.} We align the $z$-axis with this direction and then apply two criteria to select disk galaxies:
i) $k_{\rm{rot}} \geq 0.5$ and
ii) $F \leq 0.7$
Here, $k_{\rm{rot}}$ represents the fraction of rotational kinetic energy relative to total kinetic energy. Specifically, it is defined as the mass-weighted mean value of $v_{\phi}^2/v^2$ within 30 kpc, where $v$ is the total velocity and $v_{\phi}$ is the azimuthal velocity of each stellar particle \citep{sales2010}.
The morphological flatness parameter $F$ is defined as $F \equiv M_1/\sqrt{M_2, M_3}$, where $M_i$ are the eigenvalues of the moment of inertia tensor, sorted such that $M_1 \leq M_2 \leq M_3$ \citep{genel2015}.

\subsection{Selecting barred disk galaxies}
\label{barred_disks}
By computing the $m = 2$ azimuthal Fourier component of the mass distribution of the disk we define the bar strength. We project all of the particles to the $x-y$ plane, and divide the projected disk into annuli of fixed width $\Delta R=0.1$ kpc. The Fourier coefficients are given by:
\begin{eqnarray}
	a_{m} \left( R \right) &\equiv& \frac{1}{M \left( R \right)} \sum_{k=0}^{N} m_k \cos \left( m \phi_k \right), ~ {m} = 1, 2, .. \, , \\
	b_{m} \left( R \right) &\equiv& \frac{1}{M \left( R \right)} \sum_{k=0}^{N} m_k \sin \left( m \phi_k \right), ~ {m} = 1, 2, .. \, , 
\end{eqnarray}
where, $N$ represents the number of particles in the annulus, $R$ denotes the mean cylindrical radius of the annulus, $M$ is the total mass of particles within the annulus. Each particle is identified by the index $k$, with mass $m_k$ and azimuthal angle $\phi_k$. Now, the Fourier amplitude for mode $m$ at radius $R$ is defined as
\begin{eqnarray}
    A_{m} \left( R \right) ~\equiv~ \sqrt{a_{m} \left( R \right)^2 + b_{m} \left( R \right)^2}, \, \,\,\,\, m\geq 1 .
\end{eqnarray}
We note that with the definitions above we have $A_0=1$ \citep{ohta,Aguerri_2000,Athanassoula2013}. When an $\rm{m}=2$ symmetric feature, such as a bar, is present in the disk, this function typically exhibits a pronounced maximum in the internal part of the disk. We use this maximum value to define the bar strength $A_2^{\text{max}} ~\equiv~ \max [A_2(R)]$ \citep{Laurikainen2008}. Bars are typically classified into two categories: strong bars, characterized by $A_2^{\text{max}} \geq 0.4$, and weak bars, defined by $0.2 \leq A_2^{\text{max}} < 0.4$. Conversely, whereas, disks with $A_2^{\text{max}} < 0.2$ are considered unbarred \citep{algorry2017}. We take into account both the weak and strong bars in our sample.  Recalling that we select barred galaxies with stellar masses $M_* > 10^{10.0}\, M_{\odot}$, our sample consists of 589 barred galaxies, including 276 weakly barred and 313 strongly barred galaxies.

\subsection{Bar length measurement}
\label{bar-ln}

To measure the bar length we use the Fourier decomposition of the mass surface density presented in \cite{Aguerri_2000}. We first compute the intensity in the bar ($I_{\mathrm{b}}$) and inter-bar ($I_{\mathrm{ib}}$) zones, and find their ratio as a function of radius $R$ in the interval $(0.1,6)\,$kpc:
\begin{equation}
	\mathcal{I} \left( R \right) ~\equiv~ \frac{I_{\mathrm{b}} \left( R \right)}{I_{\mathrm{ib}} \left( R \right)} ~=~ \frac{A_0/2 ~+~ A_2 ~+~ A_4 ~+~ A_6}{A_0/2 ~-~ A_2 ~+~ A_4 ~-~ A_6 } \, .
	\label{Bar_interbar_ratio}
\end{equation}
The semimajor axis of the bar is defined as the outermost radius beyond which $\mathcal{I} \left( R \right)$ falls below $\left( \mathcal{I}^{\text{max}} + \mathcal{I}^{\text{min}} \right)/2$, where, $\mathcal{I}^{\text{max}}$ and $\mathcal{I}^{\text{min}}$ are the maximum and minimum values of $\mathcal{I} \left( R \right)$, respectively. The bar length is then calculated as twice the length of the semimajor axis. In some rare cases, it is necessary to extend the radial range beyond $6~\text{kpc}$ in order to measure the bar length.

\subsection{The line-of-sight velocity distribution cubes}
\label{losvd}
To apply the TW method to simulations, we first determine the position angle (PA) of the bar in the face-on view of the galaxy. We then rotate the galaxy to set the PA to $57^{\circ}$. The reference direction is the $X$-axis, and angles are measured counterclockwise. Next, we consider an observer viewing the galaxy at an inclination angle of $i = 45^{\circ}$. For this observer, the initial PA of $57^{\circ}$ appears to be approximately $45^{\circ}$. Therefore, our mock galaxies have intermediate values for the bar position angle and disk inclination angle. As a result, these galaxies are suitable candidates for pattern speed measurement using the TW method. Finally, we construct the LOSVD cubes following the procedure presented in \cite{rimoldini2014} and \cite{connor2022}. The stellar LOSVD cube for each mock galaxy has dimensions $(n_x, n_y, n_v)$, where $n_x$ and $n_y$ represent the spatial grid dimensions, and $n_v$ denotes the number of line-of-sight velocity channels. We set $n_x = n_y$, indicating a square region around the galaxy. The velocity axis spans the range $[v_{min}, v_{max}]$, with a velocity resolution of $\Delta v$ relative to the center of the galaxy. Here, $v_{max}$ and $v_{min}$ represent the maximum and minimum line-of-sight velocities of the particles in the galaxy, respectively.
We define a square region around the galaxy with side length $L$ and construct spatial pixels within this area. Assuming each pixel has a size $a$, we calculate $n_x = n_y = L/a$. Let us denote the LOSVD cube by $\mathcal{M}$. In this notation, $\mathcal{M}_k(X,Y)$ represents the $k$-th velocity channel, where $X$ and $Y$ vary across the spatial extent of the data indicating the location of the pixels. The number assigned to $\mathcal{M}_k(X,Y)$ is the total mass of the particles that are in the pixel $(X,Y)$ and have the velocities in the interval $v_{k-1}<v<v_{k+1}$. To each pixel, we assign a mass surface density $\Sigma$,\footnote{Mass-weighted quantities derived from simulations can be compared to luminosity-weighted quantities obtained from observations, by assuming an appropriate mass-to-light ratio.} a mass-weighted line-of-sight velocity $V_{\text{LOS}}$, and a mass-weighted velocity dispersion $\sigma_{\text{LOS}}$ according to the following relationships:
\begin{equation}\label{eq1}
\begin{split}
&\Sigma(X,Y)=\frac{1}{a^2}\sum_{k=1}^{n_v} \mathcal{M}_k(X,Y),\\&
V_{\text{LOS}}(X,Y)=\frac{1}{m_1} \sum_{k=1}^{n_v} \mathcal{M}_k(X,Y)\, v_k,\\&
\sigma_{\text{LOS}}^2(X,Y)=\frac{m_1}{m_1^2-m_2}\sum_{k=1}^{n_v} \mathcal{M}_k(X,Y) (v_k-V_{\text{LOS}})^2.
\end{split}
\end{equation}
Here $(X,Y)$ is the spatial coordinate of the center of the given pixel, $m_1(X,Y)=\sum_{k=1}^{n_v} \mathcal{M}_k(X,Y)$ and $m_2(X,Y)=\sum_{k=1}^{n_v} \mathcal{M}_k(X,Y)^2$, and as already noted $\mathcal{M}_k(X,Y)$ is the mass in the $k$-th velocity element ($v_k$) of a pixel. The velocity distribution function $f(X,Y,v_k)$ in the pixel $(X,Y)$ can be defined as
\begin{equation}
\label{vd}
f(X,Y,v_k)=\frac{\mathcal{M}_k(X,Y)}{V_1(X,Y)}.
\end{equation}
In all calculations presented in this paper, we set the region size to $L = 30 \,\text{kpc}$ and the velocity resolution to $\Delta v = 5 \, {\rm km \, s^{-1}}$. We  degrade the quality of the data from these high-quality baseline mock datasets and examine how effectively bar properties can be recovered. The details of the instrumental setup are provided in Section \ref{Synthetic}.

\begin{figure*}
\centering
\includegraphics[width=0.25\textwidth]{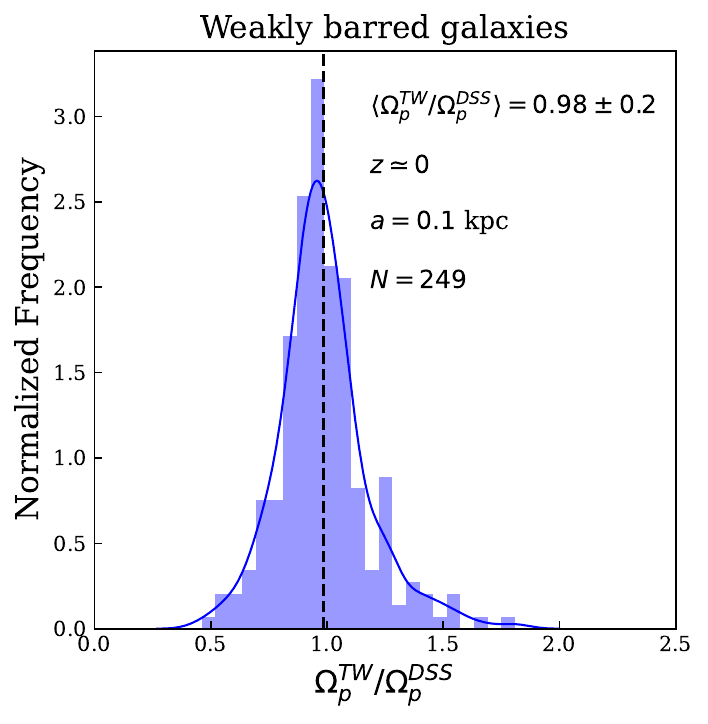}
\includegraphics[width=0.24\textwidth]{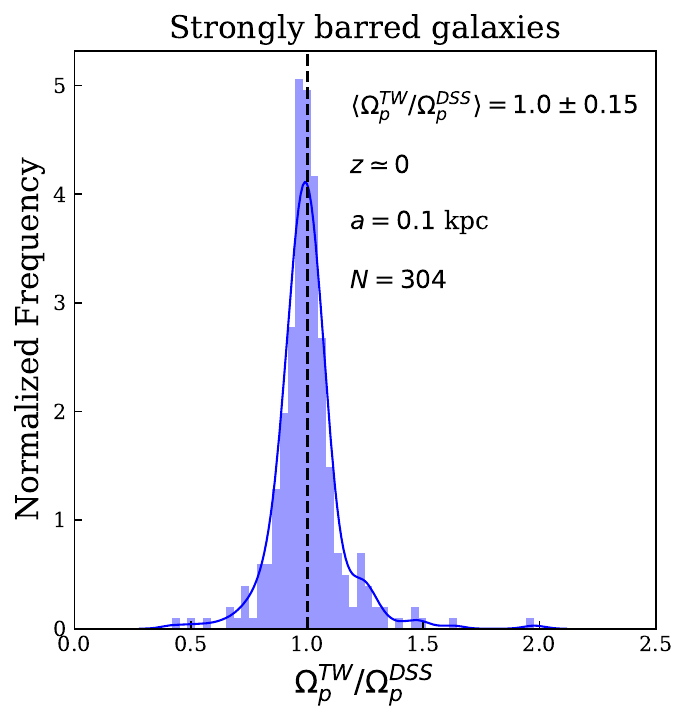}
\includegraphics[width=0.25\textwidth]{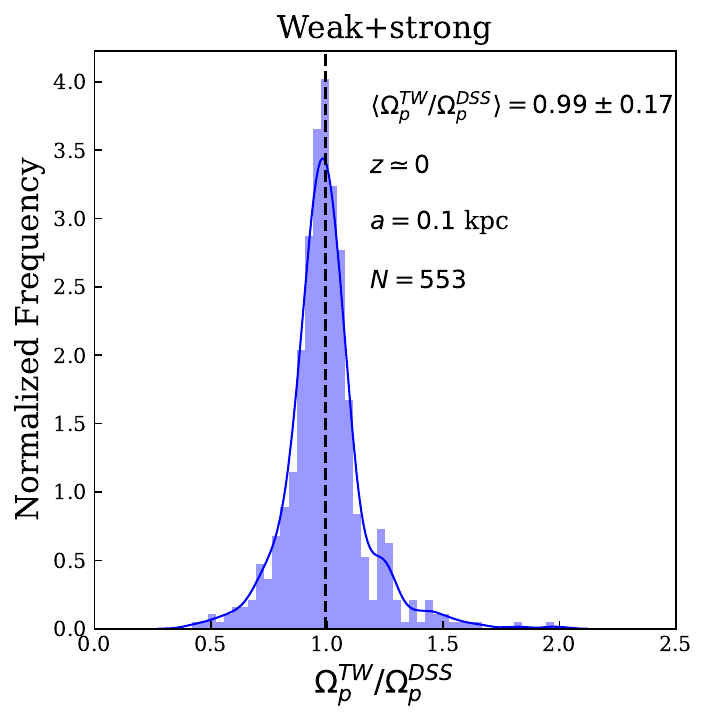}
\caption{Distribution of the ratio $\Omega^{\text{TW}}_{\rm p}/\Omega^{\text{DSS}}_{\rm p}$ (blue histogram) for the weak (left panel) and strong bars (central panel) as well as for both of them (right panel). Each panel reports the mean value (dashed vertical line), of this ratio with its corresponding error, calculated as the standard deviation around the mean. $z$ is the redshift, $a$ is the spatial sampling, and $N$ is the number of the galaxies. The blue curves in each panel represent continuous probability density functions derived from the data.}
 \label{fig2}
\end{figure*}

\section[Comparison between the TW and DSS methods]{Comparison between the TW and DSS methods}
\label{comp}

In galactic simulations, measuring $\Omega_{\rm p}$ is relatively straightforward. One can simply observe the time evolution of the bar PA and calculate the angular speed accordingly, for example see \cite{2021A&A...650L..16F} for the case of Auriga cosmological simulation. However, in this study, we deal with single snapshots of simulated galaxies, as we do not attempt to measure their time evolution. Therefore, to determine the pattern speed, we treat them as real galaxies, employing the TW method. It is important to note that the new DSS method presented in \cite{newps}, also works for single simulated snapshots. This method, which fundamentally derives from the continuity equation and utilizes an appropriate window function, is claimed to yield superior results compared to the TW method. However, the DSS method is specifically developed for simulations and is not applicable to distant galaxy observations. In simulations, we have access to all particle information and are not limited to using only line-of-sight velocities in calculations. Additionally, the window function is not limited to being solely a function of $Y$. Given these advantages, it is reasonable to expect that the DSS method provides more accurate measurements of pattern speed. It is worth mentioning that the DSS method has been implemented to measure bar pattern speed in the Milky Way and the Large Magellanic Cloud, where six-dimensional phase space data is available \citep{zhang2024,Arranz2024}.

Our objective is to investigate the performance of the TW method at high redshifts. To achieve this, it is crucial to establish the "correct" values of pattern speeds. In the Appendix \ref{test}, we conduct a test of the DSS method to verify the claims in the literature regarding its high accuracy. Therefore, throughout the remainder of the paper, we adopt the pattern speed values obtained from the DSS method as the reference values. We then compare the results of the TW method to those of the DSS method in the high-resolution case.

\subsection{The TW and DSS methods in TNG50 at high resolution}
\label{test2}

The TW method was successfully employed to measure bar pattern speeds in TNG50 simulations, as demonstrated in \cite{2021MNRAS.508..926R} and \cite{AHabibi}. Conversely, the DSS method was utilized in \cite{dehnen2024} to determine pattern speeds in the same simulations. In this study, we apply both methods to our sample of 589 barred galaxies, as described in Section \ref{Data}. We use the DSS results as a baseline and compare the TW results against them, enabling us to draw statistical conclusions about the efficacy of the TW method. It is worth noting that throughout this subsection, we operate within the high-resolution domain, with a pixel size of $a=0.1\,$kpc. We take into account more realistic conditions in subsequent sections. It is important to note that in the TW method, we adopt the standard error of the linear regression in the $\langle X \rangle-\langle V \rangle$ as the error in the pattern speed measurement. For our analysis, we retain only galaxies with relative errors smaller than 20\% in both the TW and DSS methods. This criterion ensures that we include only galaxies where both methods yield reliable results. After applying this selection, our final sample consists of 553 galaxies. Among the galaxies that leave our sample at this step, there are 29 galaxies for which the DSS method performs adequately, but the TW method produces significant errors greater than 20\%. Of these, 22 galaxies are classified as weakly barred. In observational data, weak bars are challenging for the TW method due to a weak signal or the lack of a single, well-defined pattern speed \citep{Cuomo2019a}.

Figure \ref{fig2} illustrates the ratio $\Omega^{\text{TW}}_{\rm p}/\Omega^{\text{DSS}}_{\rm p}$ for both weak and strong bars. For the 249 weak bars, the mean value of this ratio is
\begin{equation}
\langle\Omega^{\text{TW}}_{\rm p}/\Omega^{\text{DSS}}_{\rm p}\rangle=0.98\pm 0.20 ,
\end{equation}
where the error is represented by the standard deviation around the mean.
Similarly, for the 304 strong bars in our sample, we find
\begin{equation}
\langle\Omega^{\text{TW}}_{\rm p}/\Omega^{\text{DSS}}_{\rm p}\rangle=1.0\pm 0.15.
\end{equation}
By joining the weak and strong bars in the right panel in Fig. \ref{fig2}, we find
\begin{equation}
\langle\Omega^{\text{TW}}_{\rm p}/\Omega^{\text{DSS}}_{\rm p}\rangle=0.99\pm 0.17.
\end{equation}
Considering $\Omega^{\text{DSS}}_{\rm p}$ as our baseline measurement, these results demonstrate that the TW method despite its limitations performs fairly well for pattern speed measurements, especially in the case of strong bars. This comprehensive test, encompassing a large sample of simulated galaxies, reveals that the TW method introduces an error of less than 20\% for weak bars, while for strong bars the error is reduced to less than 15\%.
It is crucial to note that these results are predicated on accurate knowledge of both the inclination angle of the disk and PA of the bar. Also the spatial and spectral resolutions assumed to be high in these simulations. In the subsequent sections we relax these assumptions to mimic observations, and therefore the uncertainties in these quantities are likely to increase the error of the TW method. 

\subsection{The influence of the slit length in pattern speed measurements}
\label{slit_length}
It is worth noting that the slit length used in the TW method is $l_{\rm s}=30 \, \text{kpc}$. The results above indirectly confirm that this choice of slit length is sufficient to ensure convergence of the pattern speed to a constant value when expressed in terms of slit length. However, we conduct a test to measure the pattern speed as a function of slit length for all barred galaxies. A typical trend emerges: the pattern speed increases with slit length and eventually converges to a nearly constant value, around a slit length, different from galaxy to galaxy, not greater than 30 kpc. From an observational point of view, it may be useful to express the slit length $l_{\rm s}$, at which the pattern speed converges to a constant value, in terms of the bar length projected along the major axis of the disk $l_{\rm p}$.\footnote{In this paper, we denote the bar length by $l$, and the projected bar length along the semimajor axis in our mock galaxies by $l_{\rm p} = l \cos i \cos {\rm PA}$. Since our mock galaxies are constructed with an inclination angle of $i = 45^{\circ}$ and bar PA of $45^{\circ}$, it follows that $l_{\rm p} \approx l/2=R_{\rm bar}$. Additionally, the projected bar length measured along the minor axis of the galaxy is represented by $l_{\rm b}$. The length $l_{\rm b}$ is related to $l$ by the expression $l_{\rm b} = l \cos i \sin \mathrm{PA}$.} Similar tests have already been presented in \cite{Zou2019} and \cite{Guo2019}. However, these studies use only a single simulated galaxy and do not specify strict criteria for the convergence of the pattern speed. For an example of this test in a real galaxy, NGC 4277, we refer to \cite{Buttitta2022}. In contrast, our analysis includes 538 barred galaxies, providing a much stronger statistical foundation. Additionally, we explicitly define the convergence radius as the point beyond which the pattern speed remains nearly constant and does not change too much with further increases in slit length.
\begin{figure}
\centering
\includegraphics[width=0.33\textwidth]{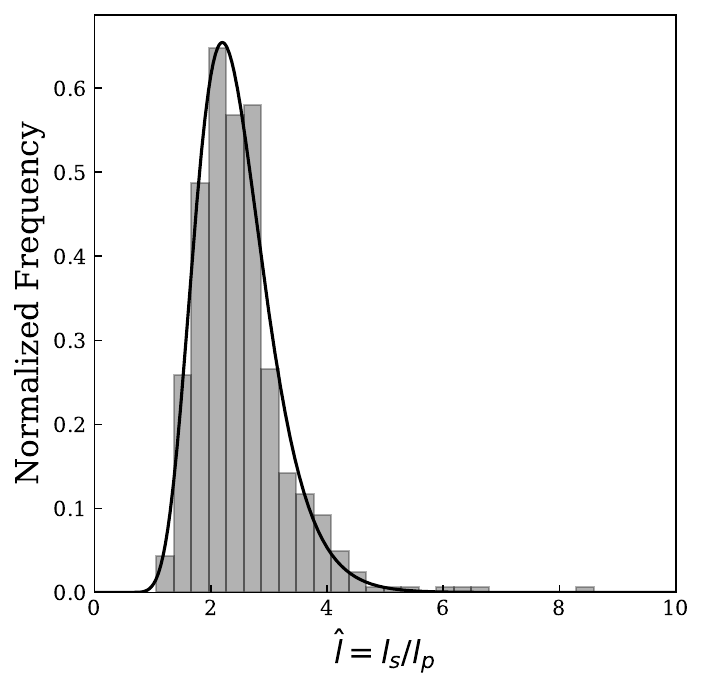}
\caption{Distribution of the ratio of the slit length $l_s$, at which the pattern speed converges to a constant value, to the projected bar length along the semimajor axis, $l_{\rm p}$. There are 538 barred galaxies (weak+strong) in the analysis. The inclination angle of all the galaxies is around $45^{\circ}$. Furthermore, the PA of the bars is around $45^{\circ}$. Therefore, $\hat{l}\simeq l_{s}/R_{\rm bar}$. The black line shows the best fit log-normal distribution function. More than 26\% of the galaxies require $\hat{l}\le 2$.}
 \label{conver}
\end{figure}

To determine the slit length $l_{\rm s}$ at which the pattern speed converges, we incrementally vary the slit length and measure the pattern speed at each step. The slit length increases by $2a = 0.2\,\mathrm{kpc}$ at each step. We calculate the mean pattern speed over the last ten steps, that is, over a slit length range of $20a = 2\,\mathrm{kpc}$. We then compute the standard error of this mean. When the standard error falls below 20\%, we stop the process and take the midpoint of that 2 kpc interval as the slit length $l_{\rm s}$ where the pattern speed converges to a constant value. We use the average pattern speed within this interval as our final measurement. Although the criteria for selecting this interval can be adjusted, our chosen method proves to be effective: using the pattern speed determined by this approach, we still obtain $\langle\Omega^{\text{TW}}{\rm p}/\Omega^{\text{DSS}}{\rm p}\rangle=0.99\pm 0.17$.

The resulting slit length differs from galaxy to galaxy. In Fig. \ref{conver}, we present the distribution of the ratio $\hat{l}=l_s/l_p$. Interestingly the distribution of $\hat{l}$ is well fit by a log-normal distribution 
\begin{equation}
p(\hat{l})\, d\hat{l}=\frac{1}{\sqrt{2\pi}\sigma}\exp\Big[-\frac{\ln^2(\hat{l}/\mu)}{2\sigma^2}\Big]\,\frac{d\hat{l}}{\hat{l}}
\end{equation}
with parameters $\mu\simeq 2.36$ and $\sigma \simeq 0.27$. Using this log-normal distribution function, the mean value  and standard deviation of $\hat{l}$ is $\langle\hat{l} \rangle\simeq 2.45\pm 0.67$.

This result may be particularly relevant in cases where spectroscopic data do not extend over a large fraction of the galaxy and  TW method is to be applied. This result may serve as an appropriate and stable guideline for planning slit lengths in observations. In our sample, we identified no galaxy with $\hat{l}\le 1$. On the other hand, there are 142 galaxies with $1<\hat{l}\le 2$. A simulated galaxy of this kind, with $\hat{l}\simeq 1.9$, has been reported by \cite{Zou2019}. On the contrary, only 20 galaxies in our sample require $\hat{l} > 4$. It may be worth noting that the most probable value of $\hat{l}$ is approximately 2.2 that covers 105 galaxies in our sample. Finally, it is important to note that the log-normal distribution may be influenced by a bias due to the prevalence of short bars in TNG50.

As a final remark, we have evaluated the performance of the TW method in high-resolution scenarios. Our findings indicate that the TW method operates with satisfactory accuracy under these conditions. This assessment now allows us to address the central question of this paper, namely how  the TW method performs at lower resolutions. It is important to note that, at least in the context of JWST observations, lower resolution typically corresponds to higher redshifts. Therefore, we can rephrase our inquiry to ask  how reliable   the TW method is when applied to galaxies at high redshifts. In the following section, we construct mock observations simulating galaxies at a redshift of $z\simeq 1.2$ as they would be observed by NIRSpec. 

\begin{figure*}
\centering
\includegraphics[width=0.216\textwidth]{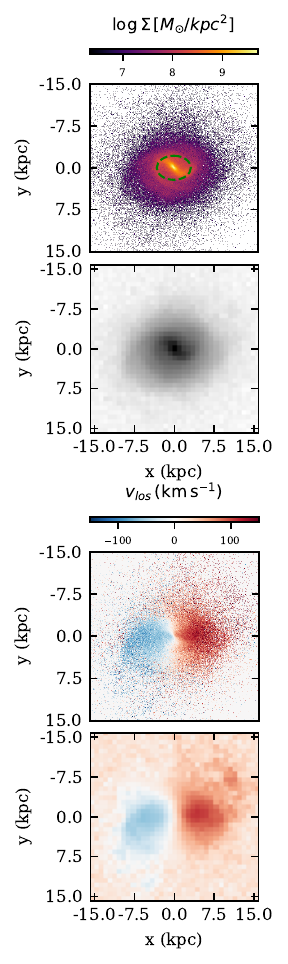}\hspace{0.01cm}
\includegraphics[width=0.17\textwidth]{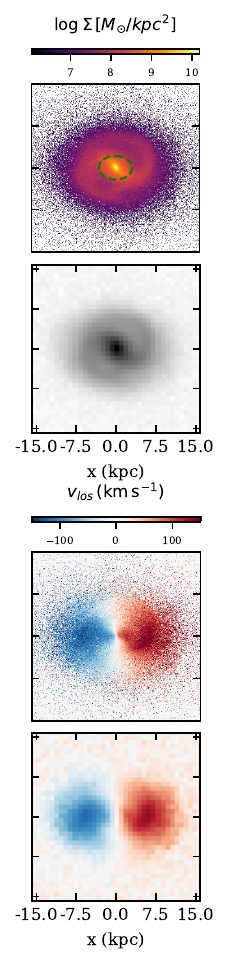}
\includegraphics[width=0.17\textwidth]{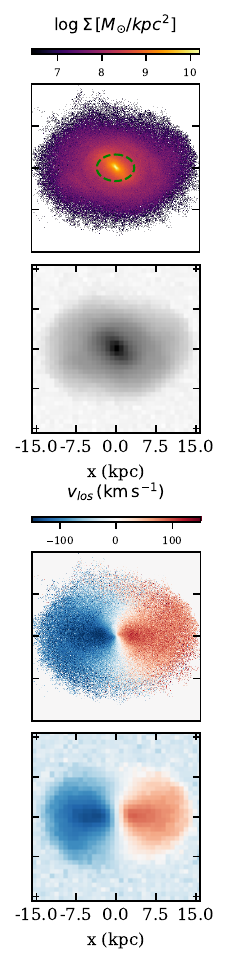}
\includegraphics[width=0.17\textwidth]{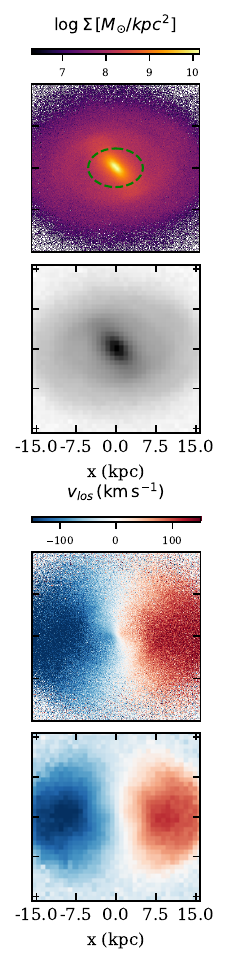}
\includegraphics[width=0.17\textwidth]{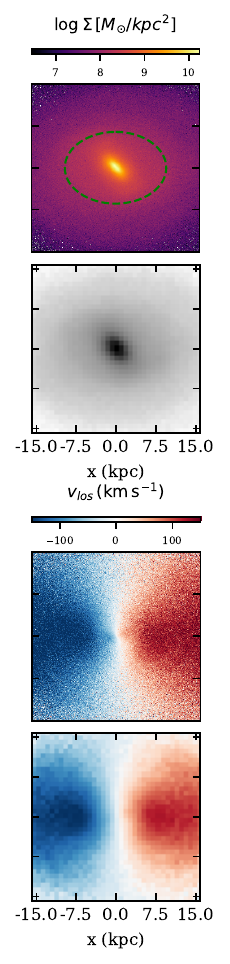}
\caption{Mass surface density and velocity maps for five TNG50 galaxies.The first row illustrates the original mass surface density of these galaxies. The second row presents their corresponding mock observations. The third row shows the original line-of-sight velocity maps, while the fourth row displays their final mock observational representations. The green ellipses in the top row indicate the half-mass radii of the galaxies.}
 \label{fig4}
\end{figure*}

\section{Synthetic JWST NIRSpec kinematic observations}
\label{Synthetic}

\subsection{Instrumental setup}
Our objective is to emulate deep IFS observations of the disk regions of a barred galaxy using NIRSpec on the JWST. We focus on the medium-resolution configuration G140M/F070LP, which offers a spectral resolution of $R\simeq 1000$ and covers the wavelength range of $0.9-1.27\, \mu$m. 
At a redshift of $z\simeq 1.2$, this configuration captures crucial absorption features in the rest-frame optical spectrum, including H$\beta$, Mg, and Fe lines. These spectral features are essential for extracting the stellar kinematics necessary for applying the TW method.
By carefully selecting an appropriate exposure time, one may achieve a signal-to-noise ratio of $S/N\simeq 5$ per pixel. This level of $S/N$ should provide sufficient data quality to perform reliable kinematic measurements while balancing observational efficiency.

NIRSpec provides a $3\,\,\!\!^{\prime\prime}\times 3\,\,\!\!^{\prime\prime}$ square field of view (FOV). This FOV is divided into 900 spatial elements, each measuring $0\rlap{\raisebox{0.5ex}{$^{\prime\prime}$}}.1\times 0\rlap{\raisebox{0.5ex}{$^{\prime\prime}$}}.1$ in size. The angular sampling of $0\rlap{\raisebox{0.5ex}{$^{\prime\prime}$}}.1 \,\text{pixel}^{-1}$ corresponds to spatial sampling of $0.854\,$kpc per pixel at the redshift $z\simeq 1.2$. Consequently, when calculating the LOSVD cubes, we set the spatial sampling parameter $a$ to $a=0.854\,$kpc. As already mentioned, we set the region size to $L = 30 \,\text{kpc}$ and the velocity resolution to $\Delta v = 5 \, {\rm km\, s^{-1}}$. After generating the cubes, it is essential to incorporate the effects of the point-spread function (PSF) and line-spread function (LSF) of NIRSpec into our data. These functions account for the instrument optical characteristics and spectral resolution, respectively. Additionally, to create more realistic simulations, we need to introduce some sources of noise to our idealized cubes. The following subsections detail the implementation of these effects, ensuring that our mock data resembles actual NIRSpec observations.

\begin{figure*}
\centering
\includegraphics[width=0.18\textwidth]{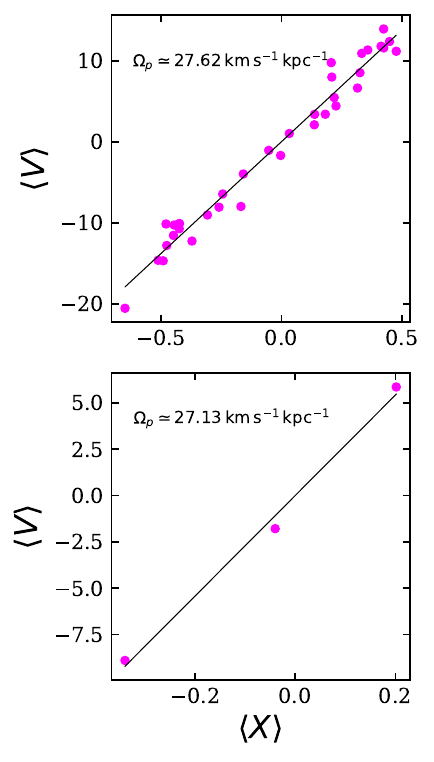}
\includegraphics[width=0.18\textwidth]{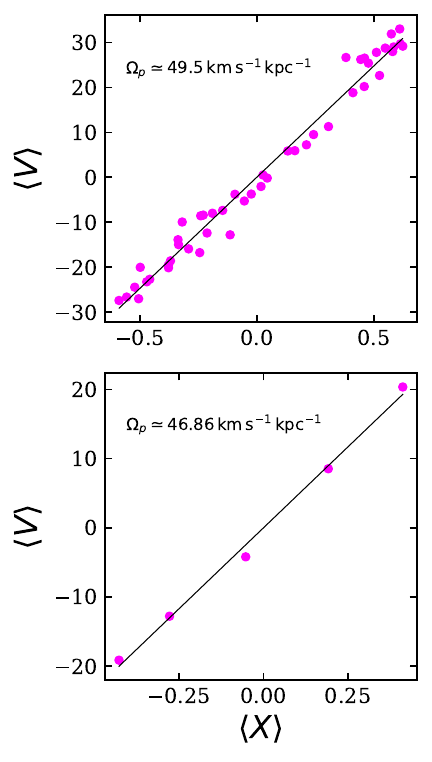}
\includegraphics[width=0.173\textwidth]{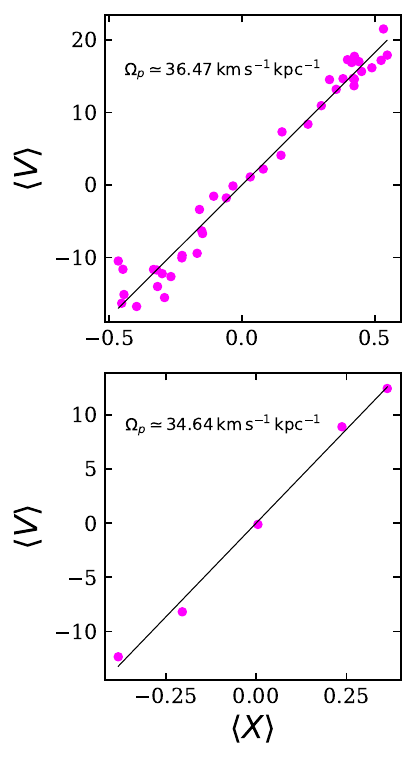}
\includegraphics[width=0.18\textwidth]{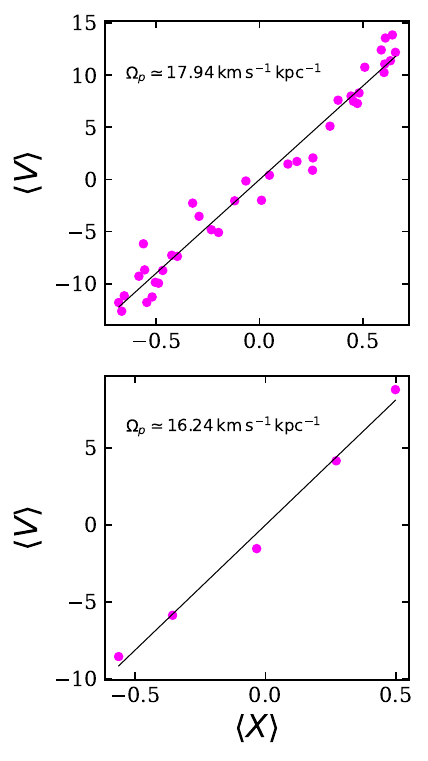}
\includegraphics[width=0.18\textwidth]{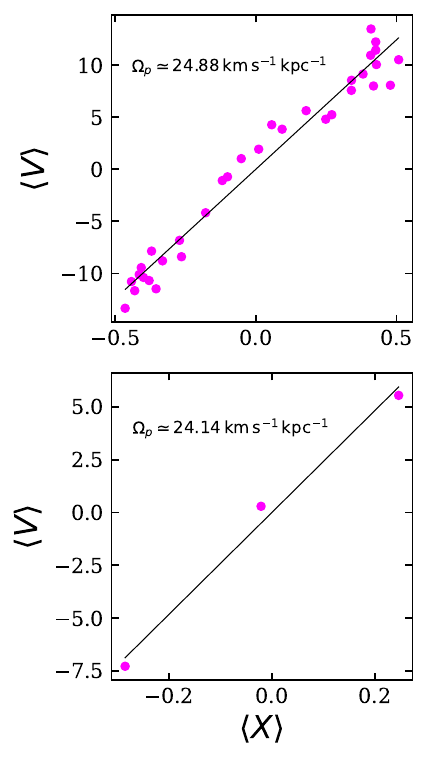}

\caption{$\langle X \rangle-\langle V \rangle$ planes for the mock galaxies presented in Fig. \ref{fig4}. The units of $\langle X \rangle$ and  $\langle V \rangle$  are kpc and ${\rm km\, s^{-1}}$,  respectively. The top row shows the high-resolution scenario with several slits, while the bottom row illustrates the high-redshift scenario with significantly fewer slits. Each purple point corresponds to a slit crossing the galaxy on the sky plane.}
 \label{fig5}
\end{figure*}

\begin{figure}
\centering
\includegraphics[width=0.33\textwidth]{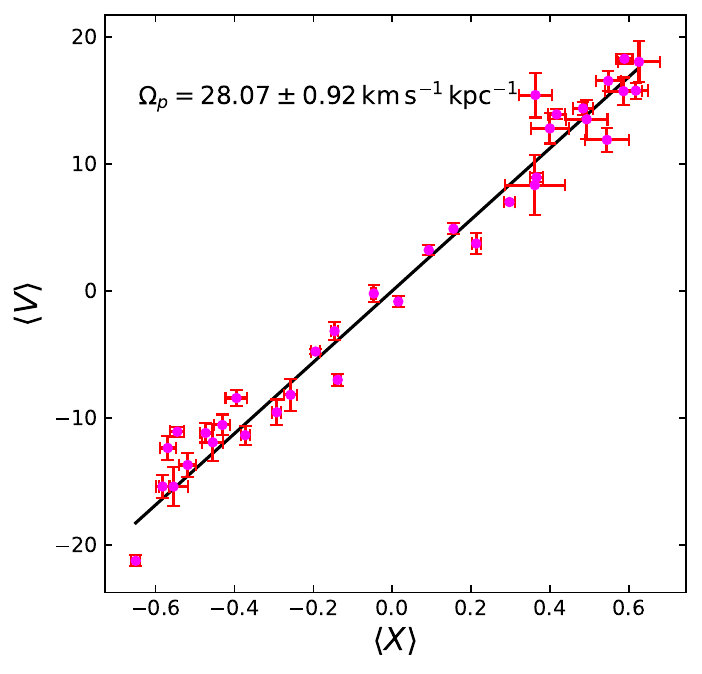}

\caption{$\langle X \rangle$–$\langle V \rangle$ plane for the galaxy with ID = 374573. This galaxy is shown in the left panel of Fig. \ref{fig4}. Unlike Fig. \ref{fig5}, the error bars of the photometric and kinematic integrals are included in the linear fit. }
 \label{fig5_new}
\end{figure}

\subsection{Point spread function}

We begin by incorporating the PSF into our cubes. For this purpose, we utilize the STPSF\footnote{\url{https://stpsf.readthedocs.io/en/latest/index.html}} software \citep{WebbPSF} to simulate the PSF of NIRSpec for the specific configuration G140M/F070LP. By selecting the aforementioned configuration and setting the pixel size to $0\rlap{\raisebox{0.5ex}{$^{\prime\prime}$}}.1\,\text{pixel}^{-1}$, STPSF generates a wavelength dependent PSF. We do not implement a PSF cube in our mocks. Instead we use an averaged PSF with a full-width at half maximum (FWHM) of $\text{FWHM}_{\text{PSF}}=0\rlap{\raisebox{0.5ex}{$^{\prime\prime}$}}.1286$, which corresponds to $\sim 1.1\,$kpc at $z\simeq 1.2$. It should be noted that among the four different PSFs generated by STPSF with varying oversampling, we opt for the output that exhibits more blurring and best represents the PSF as it would be observed on the actual detector.

We proceed by convolving each velocity channel $\mathcal{M}_k$ with the previously described PSF. Then we collapsed the convolved velocity channels to construct the PSF-convolved mass surface density map. However, before utilizing these PSF-convolved cubes for further analysis, it is essential to generate a corresponding noise cube to simulate observational noise. To simulate observational noise, we generate a spatial random Gaussian noise for each channel $\mathcal{M}_k$. The Gaussian kernel is configured with a mean value of zero, while its standard deviation $\sigma$ is carefully chosen to achieve a signal-to-noise ratio of approximately 5.
For each channel $\mathcal{M}_k$, we calculate its mean value, denoted as $\langle \mathcal{M}_k\rangle$. We then set the standard deviation $\sigma$ of the Gaussian kernel to be the ratio of this mean value to the desired $S/N$: $\sigma = \frac{\langle \mathcal{M}_k\rangle}{S/N}$. A the end, we add the noise cube to the PSF-convolved cube.

\subsection{Line spread function}

Now we are in a position to incorporate the spectral resolution into our mock observations. The spectral resolution for the instrumental setup used in this study is $R \simeq 1000$. This corresponds to a velocity resolution of $$\delta v \simeq \frac{c}{R} \simeq 300 \, \text{km} \, \text{s}^{-1}.$$ It is important to note the distinction between $\Delta v$ and $\delta v$ in our notation. This velocity resolution must be implemented through the LSF. Given that we are working with idealized LOSVD cubes and dealing with velocity channels instead of frequency channels, the velocity resolution $\delta v$ can be considered as the FWHM of the velocity distribution $f(X,Y,v)$ at each pixel, as induced by the instrument \citep{connor2022}. Assuming a Gaussian distribution for the velocity, the instrumental velocity dispersion is defined as:$$\hat{\sigma}=\frac{\text{FWHM}}{2\sqrt{2 \ln 2}}=\frac{\delta v}{2.355}= 127.4\, {\rm km\, s^{-1}}.$$ To implement this effect, we convolve a Gaussian smoothing kernel with the velocity distribution function at each pixel, using the characteristic width $\hat{\sigma}$. This process effectively simulates the spectral broadening introduced by the instrument. It turns out that despite the broadening, this smoothing procedure does not significantly alter the mean value of line-of-sight velocity in each pixel.

The resulting cubes now incorporate the spatial and spectral resolution of NIRSpec, as well as a random normal noise component. We utilize these processed cubes to generate the mass surface density and line-of-sight velocity maps, as described by the equations \eqref{eq1}. These maps are our final data set that we  use to explore the efficiency of the TW method at redshit $z\simeq 1.2$. Five mock galaxies are shown in Fig. \ref{fig4}. We include a range of galaxies, covering both compact and extended systems in terms of their effective radius. Additionally, we aim to present galaxies with three or five slits in the high redshift case. It is important to note that higher numbers of slits are rarely expected in our mock galaxies, while three and five slits appear frequently in our analysis.

\section{The TW method at high redshift $z\simeq 1.2$}
\label{highz}

We build our sample by selecting barred galaxies from three different redshifts/snapshots of the TNG50 simulation. These galaxies are treated as individual systems, each assumed to be observed at a redshift of approximately $z \simeq 1.2$. Now, we apply the TW method to such mock galaxies. Our initial sample comprised 553 galaxies, for which both the TW and DSS methods yielded accurate results in the high-resolution scenario (equivalently at $z\simeq 0$). However, the situation changes at high redshift ($z\simeq 1.2$), where our sample size significantly decreases. For the TW method to be effective, we require a minimum of three slits. This implies that the projected length, $l_{\rm b}$, of the bar along the galaxy minor axis meets a specific criterion, which can be expressed as follows:
\begin{equation}
l_{\rm b}\geq 3\times \, \text{max}[\text{FWHM}_{\text{PSF}}, a] .
\end{equation}
In our mock observations $a=0.854\,$ kpc and $\text{FWHM}_{\text{PSF}}=1.1\,$kpc. Therefore the condition simplifies to $l_{\rm b}\geq 3\times \text{FWHM}_{\text{PSF}}$. Taking into account the PA of the bar and inclination angle of the galaxy, the bar length, $l$, should satisfy the following condition to be accessible by the TW method:
\begin{equation}
l\geq \frac{3 \times \text{FWHM}_{\text{PSF}}}{\cos i\, \sin \text{PA}}\simeq 6\times \text{FWHM}_{\text{PSF}}.
\end{equation}
In our mock observations, we leveraged the fact that both the inclination angle and PA are approximately $45^{\circ}$. Our final analysis includes only those galaxies with bars longer than $ 6.6\,$kpc. It turns out that 28\% of these bars are weak and 72\% are strong. It is worth noting that bars in TNG50 appear shorter compared to observational data \citep{Frankel2022,AHabibi}. Consequently, our analysis is limited to a subset of 32 galaxies that meet these criteria. For most of these galaxies, only three slits can be implemented in the bar region. For illustration we show five mock galaxies taken from this subset in Fig. \ref{fig4}.

The inability of the TW method to measure pattern speeds of short bars at high redshifts is a limitation. However, our study focuses on galaxies accessible to the TW method, aiming to assess its accuracy/efficiency under these conditions. By choosing appropriate slits parallel to the major axis of the galaxies, we apply the TW  method. The $\langle X \rangle-\langle V \rangle$ planes for the same mock galaxies presented in Fig. \ref{fig4}  are  shown in Fig. \ref{fig5}. For comparison, the top row shows the high-resolution case with several slits, while the bottom row illustrates the high-redshift case (or, equivalently, the low-resolution case) with significantly fewer slits. Each point corresponds to a slit on the sky plane. For most of the galaxies, only three or five slits cross the bar. Since we deal with relatively short  bars, there are at most seven slits crossing the bar in our final sample of galaxies. 

It should be noted that in Fig. \ref{fig5}, the error of the linear regression is taken as the uncertainty in the measurement of $\Omega_{\rm p}$. We do not assign error bars to the integrals $\langle X \rangle$ and $\langle V \rangle$. Naturally, a more realistic approach would involve including error bars for these integrals. As previously mentioned, we defined a minimum slit length beyond which the pattern speed converges to an approximately constant value. To determine error bars for the integrals $\langle X \rangle$ and $\langle V \rangle$, we start from this minimum slit length and incrementally increase it up to twice that length. At each step, we measure the photometric and kinematic integrals. Consequently, for each slit, we obtain different values of the integrals. We then take the mean of these values as the measured $\langle X \rangle$ and $\langle V \rangle$, and their corresponding standard deviations as the errors of the integrals. Finally, we perform a linear fit that includes these error bars. The result for the left galaxy, previously shown in Fig. \ref{fig4}, is presented in Fig. \ref{fig5_new}. After applying this procedure to galaxies in Fig. \ref{fig4}, we do not expect a significant change in the results by incorporating the error bars. Therefore, we proceed with the simpler approach that ignores the error bars.

\begin{figure}
\centering
\includegraphics[width=0.33\textwidth]{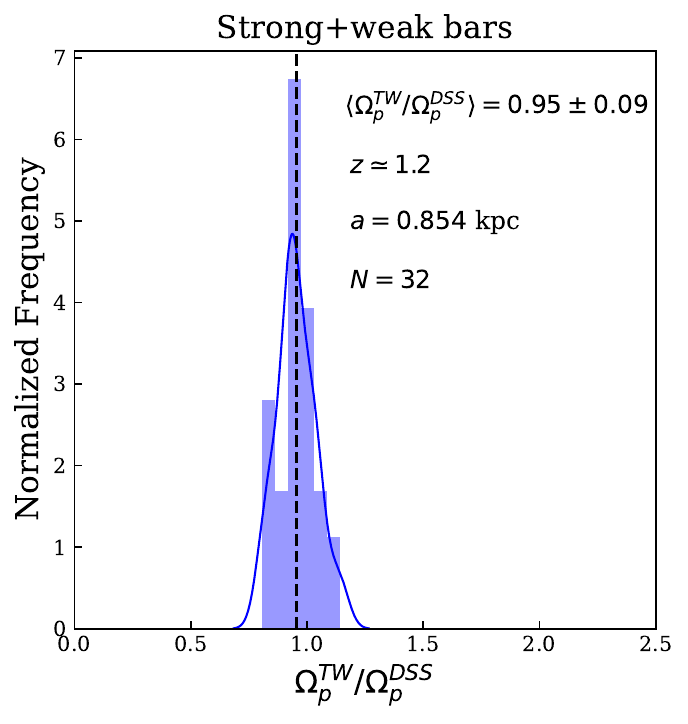}
\caption{Distribution of the ratio $\Omega^{\text{TW}}_{\rm p}/\Omega^{\text{DSS}}_{\rm p}$ (blue histogram) for 32 galaxies (9 with weak bars and 23 with strong bars) that are sufficiently extended to be detected by the TW method at $z\simeq 1.2$. Here $\Omega^{\text{TW}}_{\rm p}$ is measured at $z\simeq 1.2$, while $\Omega^{\text{DSS}}_{\rm p}$ is measured at $z\simeq 0$. The blue curve represents the continuous probability density functions derived from the data.}
 \label{fig6}
\end{figure}

After measuring the pattern speeds, we compare the results with the baseline values obtained using the DSS method at high resolution, specifically at $z\simeq 0$. The findings are illustrated in Fig. \ref{fig6}. This figure presents the distribution of the ratio $\Omega^{\text{TW}}_{\rm p}/\Omega^{\text{DSS}}_{\rm p}$ at a redshift of $z\simeq 1.2$. Our final sample includes 9 weakly barred galaxies and 23 strongly barred galaxies. The weak bars are sufficiently long to be detected by the TW method. Notably, the mean value of the aforementioned ratio is
\begin{equation}
\langle\Omega^{\text{TW}}_{\rm p}/\Omega^{\text{DSS}}_{\rm p}\rangle=0.95\pm 0.08.
\end{equation}
This result indicates that the TW method works with remarkable accuracy at $z\simeq 1.2$. In other words, this method performs exceptionally well for those galaxies hosting bars that are long enough to be accessible to the method. It is interesting to mention that the mean value of $\Omega^{\text{TW}}_{\rm p}/\Omega^{\text{DSS}}_{\rm p}$ for our 32 galaxies when mocked at $z\simeq 0$ is
 \begin{equation}
\langle\Omega^{\text{TW}}_{\rm p}/\Omega^{\text{DSS}}_{\rm p}\rangle=0.99\pm 0.09.
\end{equation}
This indicates that the uncertainties arising from limited spatial and spectral resolutions, as well as random noise, do not significantly affect the pattern speed measurements obtained using the TW method, provided its applicability conditions are satisfied. This result is consistent with the findings reported by \cite{Zou2019}.

\begin{figure}
\centering
\includegraphics[width=0.35\textwidth]{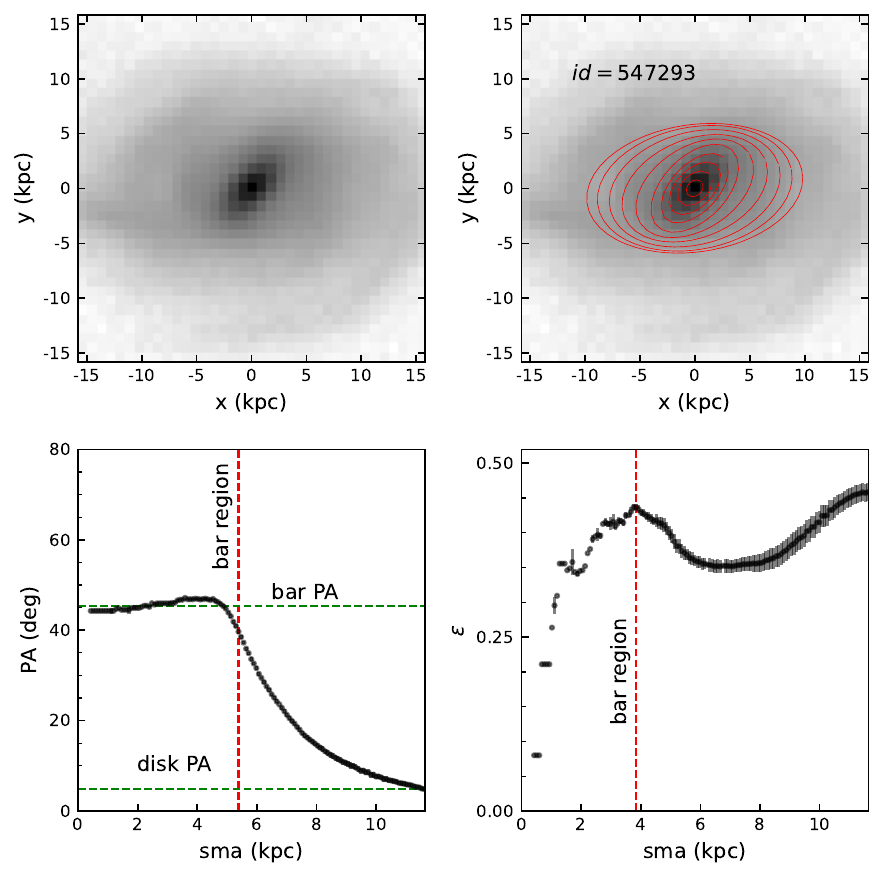}
\caption{Isophotal analysis for the mock galaxy with ID= 547293. The top right panel displays the original mock galaxy, while the top left panel illustrates the ellipse fitting process. In the bottom right panel, we present the ellipticity ($\epsilon$) as a function of the semimajor axis of the ellipses;  the red dashed line marks the projected bar length. The horizontal green dashed lines show the PA of the bar and disk, which is close to zero.The bottom left panel shows the PA of the ellipses as a function of the semimajor axis of the ellipses. Here the red dashed line indicates the inner bar region adopted to calculate the mean PA of the bar.}
 \label{fig7}
\end{figure}
\begin{figure}
\centering
\includegraphics[width=0.35\textwidth]{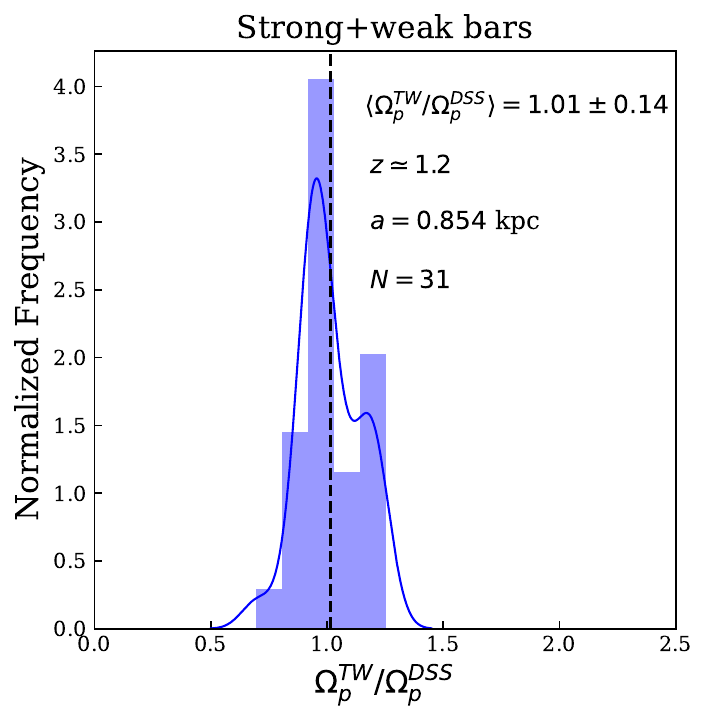}
\caption{Distribution of the ratio $\Omega^{\text{TW}}_{\rm p}/\Omega^{\text{DSS}}_{\rm p}$ (blue histogram) for 31 galaxies (comprising  weak and strong bars) that are sufficiently extended to be detected by the TW method at $z\simeq 1.2$. The bar PA and the inclination of the disks have been measured using the isophotal analysis. The blue curve represents the continuous probability density functions derived from the data.}
 \label{fig8}
\end{figure}
\section{ PA and inclination angle measurement}
\label{extra}
As already mentioned, the results presented in the previous sections are based on our precise knowledge of the position angle of the bar, as well as inclination angle of the galaxy. In this section, we conduct an additional test to assess the robustness of the TW method against these properties. We treat our mock galaxies as if they were real observational targets, employing a widely used technique to measure their bar position angle and disk inclination\citep [e.g.,][] {mobasher2023,liang}. This approach mirrors the process used in actual galactic observations, providing a more realistic situation for our test. For this analysis, we utilize the 32 galaxies of our final sample. We use isophotal ellipse fitting method using the {\sc photutils} from Python {\sc astropy} package. This method is based on the radial profiles of the ellipticity $\epsilon$ and PA as a function of the semimajor axis (sma) of the ellipses that best fit the galaxy isophotes. We extend the outermost ellipse to the region where the $S/N$ falls to a value of 10 ~pixel$^{-1}$ or less.\footnote{We also conducted the analysis using a $S/N$ threshold of 3~pixel$^{-1}$ . Although the overall results did not change substantially, the uncertainty in the findings increased.} We initially allowed the center of the ellipses to vary. Subsequently, we used the inner ellipses to determine the center and then reran the {\sc photutils} with the same fixed center applied to all ellipses.

Given the presence of prominent bars in our galaxies, the PA typically remains constant within the bar region. We calculate the mean PA value in the inner bar region and then determine the radius, $a_1$, where the PA deviates by $\Delta\theta=5^{\circ}$ from this mean. This radius is marked by the red dashed line in the bottom left panel of Fig. \ref{fig7} for one of our mock galaxies. We consider the mean PA value, calculated up to $a_1$, to be representative of the bar PA. Since we are going to repeat the TW analysis, we need bar PA in order to determine the number of slits. For now, we do not measure the PA of the disk. Instead, we utilize the fact that the semimajor axis of all our mock galaxies is aligned with the $x$-axis, which implies that the PA of the disk is zero. In this manner, we can quantify the impact of the bar PA and disk inclination on the accuracy of the TW method. Finally, at the end of this section, we relax our assumption about knowing the disk PA and instead measure it using both isophotal and kinematic analysis.

To determine the number of slits, we also need to establish the projected bar length along the galaxy minor axis. In the bar-dominated region, we expect the ellipticity $\epsilon$ to increase smoothly until it reaches a maximum value exceeding $0.25$. We denote this radius as $a_2$. Selecting $\text{min}(a_1,a_2)$ as the projected bar length represents a conservative choice for the value for the bar length. While other criteria for determining bar length exist in the literature \citep[e.g.,][]{mobasher2023}, we opt for this conservative method to ensure that all slits are positioned within the bar region. As an illustration, we have indicated $a_2$ with a red dashed vertical line in the bottom right panel of Fig. \ref{fig7} for one of our mock galaxies.

As for the disk inclination angle, we take the ellipticity of the outermost ellipse. Then the inclination angle is obtained by
\begin{equation}
i=\arccos (1-\epsilon).
\end{equation}
By assuming an infinitesimally thin disk, we use this inclination along with the number of slits determined from the bar position angle and projected length, we recalculate the pattern speeds employing the TW method. We reject galaxies where the relative error in pattern speed exceeds 20\%. As a result, one galaxy is removed from the sample. The result of this analysis is presented in Fig. \ref{fig8}. The mean value of $\Omega^{\text{TW}}_{\rm p}/\Omega^{\text{DSS}}_{\rm p}$ is
 \begin{equation}
\langle\Omega^{\text{TW}}_{\rm p}/\Omega^{\text{DSS}}_{\rm p}\rangle=1.01\pm 0.13. 
\end{equation}
The mean value of $\hat{i} = \frac{i}{45^{\circ}}$ and corresponding standard deviation obtained using this method is $\langle\hat{i}\rangle = 0.93 \pm 0.12$. This indicates that isophotal analysis nicely performs for measuring inclination angles. As anticipated, uncertainties in the bar position angle and galaxy inclination angle lead to increased, although small, uncertainty in the pattern speed measurements. Our analysis reveals that the error in pattern speed can reach up to approximately 14\%. Despite this minor increase in uncertainty, the results remain reasonably acceptable, demonstrating that the TW method maintains its effectiveness and reliability under these more realistic conditions. 

However, it is well established that uncertainty in the PA of the disk is the primary source of error for this method \citep{deb2003}. For another test, we determine the PA of the disk using isophotal and kinematic analysis. In the isophotal analysis, the PA of the outermost ellipse serves as an indicator of the disk PA. Some of the galaxies are large in size, and the mock box with size of $L=30\,$kpc does not fully encompass them. For these galaxies,\footnote{The ID numbers are 314577, 307487, 396628, 184932, and 471248. The galaxy with ID=184932 is shown in the fourth column from the left in Fig. \ref{fig4}.} the isophotal analysis does not work. Furthermore, we have four spiral galaxies\footnotetext{The ID numbers are 318883, 415186, 374573, and 547293. The galaxies 374573 and 318883 are shown in the first and second columns of Fig. \ref{fig4}, respectively.} for which the isophotal analysis is challenging. For these two subset of galaxies, we replace the photometric PA with the kinematic PA obtained from the velocity map. The kinematic PA was measured using the {\sc PaFit} software package \citep{pafit}, which applies the method presented in \cite{pafit2} to determine the orientation of the galaxy's rotation axis.

Although the position angles should be close to zero for all mock galaxies, the isophotal analysis does not precisely meet this expectation. Instead, the mean position angle and its corresponding standard deviation are $\langle\delta_{\rm PA}\rangle=-0\rlap{\raisebox{0.5ex}{$^\circ$}}.99\pm 3\rlap{\raisebox{0.5ex}{$^\circ$}}.32$. This level of uncertainty, as expected \citep{deb2003}, leads to a significant error in the method:
 \begin{equation}
\langle\Omega^{\text{TW}}_{\rm p}/\Omega^{\text{DSS}}_{\rm p}\rangle=1.18\pm 0.47.
\label{pa_ph}
\end{equation}

As previously done, we exclude galaxies with a relative error in pattern speed exceeding 20\%. Additionally, we identify galaxies where large errors in the disk PA measurement cause the sign of the pattern speed to differ from the baseline value. These galaxies are also excluded, because the sign of the pattern speed is clearly evident from the velocity maps of the galaxies, making it straightforward to recognize incorrect rotation verse. More specifically, two galaxies are removed for the wrong sign of the pattern speed and three for the error in the magnitude of the pattern speed. Therefore, the final sample includes 27 galaxies.

It is worth noting that galaxies with $\delta_{\rm PA} > 0$ exhibit $\Delta\Omega = \Omega_p^{\rm TW} - \Omega_p^{\rm DSS} < 0$, whereas galaxies with $\delta_{\rm PA} < 0$ show $\Delta\Omega > 0$. This relationship holds true for all galaxies in our sample and is consistent with the findings reported in \cite{deb2003}. However, it is important to point out that the definition of $\delta_{\rm PA}$ in our paper differs in sign from that in \cite{deb2003}. Specifically, we align the major axis of the galaxy along the $x$-axis before measuring $\delta_{\rm PA}$, whereas \cite{deb2003} aligns the slits along the $x$-axis. 

We confirm previous findings \citetext{see, e.g., \citealp{deb2003} and \citealp{Zou2019} }, that uncertainty in the PA of the disk is the primary source of error in the TW method. Regardless of whether galaxies are at low or high redshifts, accurate measurement of the disk PA is a critical step of the TW method. Consequently, at high redshifts, the reduced spectral and spatial resolution does not affect the effectiveness of the TW method. Additionally, uncertainties in the PA of the bar and inclination of the disk do not significantly change the results. Nonetheless, it is essential to exercise caution when measuring the disk PA.

It is important to emphasize that, in practice, the error in the PA measurement can be further reduced. Specifically, our mock data cubes simulate the spectroscopic data from NIRSpec, whereas the photometric data obtained by NIRCam offers higher resolution for isophotal analysis. In other words, the PA of the disk can be measured more precisely using the higher-resolution photometric data. This improved measurement reduces the uncertainty in the disk PA and, consequently, decreases the error in the pattern speed. As the final remark, this paper focuses on the F070LP filter in NIRSpec. For qualitative discussions of other filters, see Appendix \ref{JWST_TW}.

\section{Summary and conclusion}
\label{discussion}
The TW method has not yet been applied to high-redshift observations. This paper explores the functionality of the method at redshift $z\simeq 1.2$. To achieve this, we constructed mock observations emulating JWST's NIRSpec observations using barred galaxies from the TNG50 cosmological simulation. To assess the method's efficiency, it was necessary to establish true values for the pattern speeds of our mock galaxies. We employed the DSS method presented in \cite{newps} to measure baseline values for the pattern speed. This method is exclusively applicable to simulated data and cannot be directly implemented in real observational scenarios.

We conducted a brief validation of the DSS method. We compared results obtained from this method with the exact pattern speed evolution of an isolated galaxy simulated using the {\small GALAXY} code. Our findings reaffirmed the claim in the literature that the DSS method operates with high accuracy. Before progressing to high redshifts, or equivalently low-resolution scenarios, it was essential to evaluate the efficiency of the TW method against the DSS method in the high-resolution case. We provided this assessment in section \ref{test2}, verifying that the TW method functions efficiently in high-resolution scenarios. Specifically, our analysis revealed that for 249 weakly barred galaxies in our sample, the TW method resulted in errors of approximately 20\%. Conversely, for 304 strongly barred galaxies, the error was less than 15\%. Overall, considering both types of barred galaxies, the TW method operated with an average error of 17\%. Given that we examined a large sample of galaxies with diverse characteristics, our results lend considerable credibility to the efficiency of the TW method.

We then constructed mock observations emulating NIRSpec observations in the medium-resolution configuration G140M/F070LP, which provides a spectral resolution of $R\simeq 1000$. Rather than performing a full radiative transfer analysis on our mock galaxies, we opted to construct LOSVD cubes. To create our final mock galaxies, we convolved the velocity channels with appropriate PSF and LSF, and incorporated random normal noise into the cubes to achieve a signal-to-noise of $S/N\sim 5$. The PSF for NIRSPec at this specific configuration is simulated by the STPSF software. For consistency across all galaxies in our sample, we set the position angle of the bar and the inclination angle of the galaxy to $45^{\circ}$. 

Given that the spatial size of pixels in our final images is $a=0.854\,$kpc, we discovered that the TW method is inapplicable for 94\% of our galaxies. This limitation arises because the bar lengths in this portion of galaxies are too short to accommodate at least three slits within the bar region, which is a requirement for the TW method. Consequently, we can conclude that at high redshifts, the applicability of the TW method is constrained by the size of the ${\rm FWHM_{PSF}}$ or the spatial size of the pixels, effectively limiting its use to galaxies with bars exceeding a minimum required length. In the context of our instrumental setup for this study, this minimum bar length is approximately $\sim 6.6\,$kpc. The next question we addressed is  how efficient   the TW method is for those galaxies that are accessible to it at high redshift. Given the aforementioned restrictions, 32 galaxies remained in our sample that were suitable for analysis using the TW method. For these mock galaxies, we applied the TW method using the exact values of the bar's PA, the galaxy's PA, and the galaxy's inclination angle. Notably, we found that the method performs with high accuracy under these conditions, with the uncertainty in pattern speed measurements remaining below 10\%.

In the final stage of our analysis, we treated our mock galaxies as if they were real observational targets, measuring their inclination angles and bar PAs using isophotal analysis. At this step, we used the exact and already known values for the disks' PA. This approach introduces uncertainties in these parameters, which in turn slightly affect the accuracy of pattern speed measurements via the TW method. As anticipated, this resulted in a bit higher error margin of approximately $\sim 15\%$ in the TW method's performance. Then, we also determined the disk PA using isophotal and kinematic analysis. This method yields a mean uncertainty of $\langle\delta_{\rm PA}\rangle=-0\rlap{\raisebox{0.5ex}{$^\circ$}}.99\pm 3\rlap{\raisebox{0.5ex}{$^\circ$}}.32$ in PA measurements. This level of uncertainty leads to errors in the TW method as $\langle\Omega^{\text{TW}}_{\rm p}/\Omega^{\text{DSS}}_{\rm p}\rangle=1.18\pm 0.47$, which is still acceptable from an observational point of view. 

Therefore, the key conclusion of this study is that at high redshift ($z\simeq 1.2$) using kinematic observations from JWST's NIRSpec, the TW method can be expected to function reliably for galaxies that host sufficiently long bars (see Table \ref{tab1}) and intermediate values for PAs and inclination angles ($\simeq 45^{\circ}$). 
\begin{acknowledgements}
MR is grateful to Tahere Kashfi for the discussions on the test provided in Appendix \ref{test}. The work of MR and AH is supported by the Ferdowsi University of Mashhad. This work is based upon research funded by Iran National Science Foundation (INSF) under project No.4030079. JALA acknowledge support from the Agencia Estatal de Investigación del Ministerio de Ciencia, Innovación y Universidades (MCIU/AEI) under grant ``WEAVE: EXPLORING THE COSMIC ORIGINAL SYMPHONY, FROM STARS TO GALAXY CLUSTERS” and the European Regional Development Fund (ERDF) with reference PID2023-153342NB-I00 / 10.13039/501100011033. VC acknowledges the support provided by ANID through FONDECYT research grants no. 3220206 and no. 11250723. CB gratefully acknowledges support from the Forrest Research Foundation. The project that gave rise to these results received the support of a fellowship from the “la Caixa” Foundation (ID 100010434). The fellowship code is LCF/BQ/PR24/12050015. LC acknowledges support from grants PID2022-139567NB-I00 and PIB2021-127718NB-I00 funded by the Spanish Ministry of Science and Innovation/State Agency of Research  MCIN/AEI/10.13039/501100011033 and by “ERDF A way of making Europe”. EMC and AP acknowledge the support by the Italian Ministry for Education University and Research (MIUR) grant PRIN 2022 2022383WFT “SUNRISE", CUP C53D23000850006 and Padua University grants DOR 2022-2024. TK acknowledges support from the Basic Science Research Program through the National Research Foundation of Korea (NRF) funded by the Ministry of Education (No. RS-2023-00240212 and  No. 2022R1A4A3031306). Y.H.L. acknowledges support from the Basic Science Research Program through the National Research Foundation of Korea (NRF), funded by the Ministry of Education (No. RS-2023-00249435) and the Korean government (MSIT; No. 2022R1A4A3031306). JMA acknowledges the support of the Agencia Estatal de Investigación del Ministerio de Ciencia e Innovación (MCIN/AEI/10.13039/501100011033) under grant nos. PID2021-128131NB-I00 and CNS2022-135482 and the European Regional Development Fund (ERDF) ‘A way of making Europe’ and the ‘NextGenerationEU/PRTR’. AdLC  acknowledges financial support from the Spanish Ministry of Science and Innovation (MICINN) through RYC2022-035838-I and PID2021-128131NB-I00 (CoBEARD project).
\end{acknowledgements}

\bibliographystyle{aa}
\bibliography{aa55547}

\begin{thebibliography}{100}
\expandafter\ifx\csname natexlab\endcsname\relax\def\natexlab#1{#1}\fi

\bibitem[{{Aguerri} {et~al.}(2003){Aguerri}, {Debattista}, \&
  {Corsini}}]{Aguerri2003}
{Aguerri}, J.~A.~L., {Debattista}, V.~P., \& {Corsini}, E.~M. 2003, \mnras,
  338, 465

\bibitem[{{Aguerri} {et~al.}(2001){Aguerri}, {Hunter}, {Prieto}, {Varela},
  {Gottesman}, \& {Mu{\~n}oz-Tu{\~n}{\'o}n}}]{Aguerri2001}
{Aguerri}, J.~A.~L., {Hunter}, J.~H., {Prieto}, M., {et~al.} 2001, \aap, 373,
  786

\bibitem[{{Aguerri} {et~al.}(2015){Aguerri}, {M{\'e}ndez-Abreu},
  {Falc{\'o}n-Barroso}, {Amorin}, {Barrera-Ballesteros}, {Cid Fernandes},
  {Garc{\'\i}a-Benito}, {Garc{\'\i}a-Lorenzo}, {Gonz{\'a}lez Delgado},
  {Husemann}, {Kalinova}, {Lyubenova}, {Marino}, {M{\'a}rquez}, {Mast},
  {P{\'e}rez}, {S{\'a}nchez}, {van de Ven}, {Walcher}, {Backsmann},
  {Cortijo-Ferrero}, {Bland-Hawthorn}, {del Olmo}, {Iglesias-P{\'a}ramo},
  {P{\'e}rez}, {S{\'a}nchez-Bl{\'a}zquez}, {Wisotzki}, \&
  {Ziegler}}]{Aguerri2015}
{Aguerri}, J.~A.~L., {M{\'e}ndez-Abreu}, J., {Falc{\'o}n-Barroso}, J., {et~al.}
  2015, \aap, 576, A102

\bibitem[{{Aguerri} {et~al.}(2000){Aguerri}, {Mu{\~n}oz-Tu{\~n}{\'o}n},
  {Varela}, \& {Prieto}}]{Aguerri_2000}
{Aguerri}, J.~A.~L., {Mu{\~n}oz-Tu{\~n}{\'o}n}, C., {Varela}, A.~M., \&
  {Prieto}, M. 2000, A\&A, 361, 841

\bibitem[{{Algorry} {et~al.}(2017){Algorry}, {Navarro}, {Abadi}, {Sales},
  {Bower}, {Crain}, {Dalla Vecchia}, {Frenk}, {Schaller}, {Schaye}, \&
  {Theuns}}]{algorry2017}
{Algorry}, D.~G., {Navarro}, J.~F., {Abadi}, M.~G., {et~al.} 2017, \mnras, 469,
  1054

\bibitem[{{Amvrosiadis} {et~al.}(2025{\natexlab{a}}){Amvrosiadis}, {Lange},
  {Nightingale}, {He}, {Frenk}, {Oman}, {Smail}, {Swinbank}, {Fragkoudi},
  {Gadotti}, {Cole}, {Borsato}, {Robertson}, {Massey}, {Cao}, \&
  {Li}}]{2025MNRAS.537.1163A}
{Amvrosiadis}, A., {Lange}, S., {Nightingale}, J.~W., {et~al.}
  2025{\natexlab{a}}, \mnras, 537, 1163

\bibitem[{{Amvrosiadis} {et~al.}(2025{\natexlab{b}}){Amvrosiadis}, {Wardlow},
  {Birkin}, {Smail}, {Swinbank}, {Nightingale}, {Bertoldi}, {Brandt}, {Casey},
  {Chapman}, {Chen}, {Cox}, {da Cunha}, {Dannerbauer},
  {Dudzevi{\v{c}}i{\={u}}t{\.{e}}}, {Gullberg}, {Hodge}, {Knudsen}, {Menten},
  {Walter}, \& {van der Werf}}]{2025MNRAS.536.3757A}
{Amvrosiadis}, A., {Wardlow}, J.~L., {Birkin}, J.~E., {et~al.}
  2025{\natexlab{b}}, \mnras, 536, 3757

\bibitem[{{Athanassoula}(1992)}]{at1992}
{Athanassoula}, E. 1992, \mnras, 259, 328

\bibitem[{{Athanassoula} {et~al.}(2013){Athanassoula}, {Machado}, \&
  {Rodionov}}]{Athanassoula2013}
{Athanassoula}, E., {Machado}, R. E.~G., \& {Rodionov}, S.~A. 2013, \mnras,
  429, 1949

\bibitem[{{Baes} {et~al.}(2024){Baes}, {Gebek}, {Tr{\v{c}}ka}, {Camps}, {van
  der Wel}, {Abdurro'uf}, {Andreadis}, {Tulu}, {Emana}, {Fritz}, {Kelly},
  {Kova{\v{c}}i{\'c}}, {La Marca}, {Martorano}, {Mosenkov}, {Nersesian},
  {Rodriguez-Gomez}, {Tortora}, {Vander Meulen}, \& {Wang}}]{Baes2024}
{Baes}, M., {Gebek}, A., {Tr{\v{c}}ka}, A., {et~al.} 2024, \aap, 683, A181

\bibitem[{{Barrientos Acevedo} {et~al.}(2023){Barrientos Acevedo}, {van der
  Wel}, {Baes}, {Grand}, {Kapoor}, {Camps}, {de Graaff}, {Straatman}, \&
  {Bezanson}}]{2023MNRAS.524..907B}
{Barrientos Acevedo}, D., {van der Wel}, A., {Baes}, M., {et~al.} 2023, \mnras,
  524, 907

\bibitem[{{Beckman} {et~al.}(2018){Beckman}, {Font}, {Borlaff}, \&
  {Garc{\'\i}a-Lorenzo}}]{Beckman2018}
{Beckman}, J.~E., {Font}, J., {Borlaff}, A., \& {Garc{\'\i}a-Lorenzo}, B. 2018,
  \apj, 854, 182

\bibitem[{{Bottrell} \& {Hani}(2022)}]{connor2022}
{Bottrell}, C. \& {Hani}, M.~H. 2022, \mnras, 514, 2821

\bibitem[{{Buta}(1986)}]{Buta1986}
{Buta}, R. 1986, \apjs, 61, 609

\bibitem[{{Buta}(1995{\natexlab{a}})}]{Buta1995a}
{Buta}, R. 1995{\natexlab{a}}, \apjs, 96, 39

\bibitem[{{Buta}(1995{\natexlab{b}})}]{buta1995}
{Buta}, R. 1995{\natexlab{b}}, Astrophysical Letters and Communications, 31, 17

\bibitem[{{Buttitta} {et~al.}(2022){Buttitta}, {Corsini}, {Cuomo}, {Aguerri},
  {Coccato}, {Costantin}, {Dalla Bont{\`a}}, {Debattista}, {Iodice},
  {M{\'e}ndez-Abreu}, {Morelli}, \& {Pizzella}}]{Buttitta2022}
{Buttitta}, C., {Corsini}, E.~M., {Cuomo}, V., {et~al.} 2022, \aap, 664, L10

\bibitem[{{Buttitta} {et~al.}(2024){Buttitta}, {Corsini}, \&
  {Debattista}}]{Buttitta2024}
{Buttitta}, C., {Corsini}, E.~M., \& {Debattista}, V.~P. 2024, in EAS2024,
  European Astronomical Society Annual Meeting, 457

\bibitem[{{Camps} {et~al.}(2016){Camps}, {Trayford}, {Baes}, {Theuns},
  {Schaller}, \& {Schaye}}]{Camps2016}
{Camps}, P., {Trayford}, J.~W., {Baes}, M., {et~al.} 2016, \mnras, 462, 1057

\bibitem[{{Cappellari} {et~al.}(2007){Cappellari}, {Emsellem}, {Bacon},
  {Bureau}, {Davies}, {de Zeeuw}, {Falc{\'o}n-Barroso}, {Krajnovi{\'c}},
  {Kuntschner}, {McDermid}, {Peletier}, {Sarzi}, {van den Bosch}, \& {van de
  Ven}}]{pafit}
{Cappellari}, M., {Emsellem}, E., {Bacon}, R., {et~al.} 2007, \mnras, 379, 418

\bibitem[{{Corsini}(2011)}]{corsini2011}
{Corsini}, E.~M. 2011, Memorie della Societa Astronomica Italiana Supplementi,
  18, 23

\bibitem[{{Corsini} {et~al.}(2005){Corsini}, {Debattista}, \&
  {Aguerri}}]{Corsini2005}
{Corsini}, E.~M., {Debattista}, V.~P., \& {Aguerri}, J.~A.~L. 2005, in
  Multiwavelength Mapping of Galaxy Formation and Evolution, ed. A.~{Renzini}
  \& R.~{Bender}, 370--371

\bibitem[{{Costantin} {et~al.}(2023{\natexlab{a}}){Costantin},
  {P{\'e}rez-Gonz{\'a}lez}, {Guo}, {Buttitta}, {Jogee}, {Bagley}, {Barro},
  {Kartaltepe}, {Koekemoer}, {Cabello}, {Corsini}, {M{\'e}ndez-Abreu}, {de la
  Vega}, {Iyer}, {Bisigello}, {Cheng}, {Morelli}, {Arrabal Haro}, {Buitrago},
  {Cooper}, {Dekel}, {Dickinson}, {Finkelstein}, {Giavalisco}, {Holwerda},
  {Huertas-Company}, {Lucas}, {Papovich}, {Pirzkal}, {Seill{\'e}},
  {Vega-Ferrero}, {Wuyts}, \& {Yung}}]{2023Natur.623..499C}
{Costantin}, L., {P{\'e}rez-Gonz{\'a}lez}, P.~G., {Guo}, Y., {et~al.}
  2023{\natexlab{a}}, \nat, 623, 499

\bibitem[{{Costantin} {et~al.}(2023{\natexlab{b}}){Costantin},
  {P{\'e}rez-Gonz{\'a}lez}, {Vega-Ferrero}, {Huertas-Company}, {Bisigello},
  {Buitrago}, {Bagley}, {Cleri}, {Cooper}, {Finkelstein}, {Holwerda},
  {Kartaltepe}, {Koekemoer}, {Nelson}, {Papovich}, {Pillepich}, {Pirzkal},
  {Tacchella}, \& {Yung}}]{2023ApJ...946...71C}
{Costantin}, L., {P{\'e}rez-Gonz{\'a}lez}, P.~G., {Vega-Ferrero}, J., {et~al.}
  2023{\natexlab{b}}, \apj, 946, 71

\bibitem[{{Cuomo} {et~al.}(2020){Cuomo}, {Aguerri}, {Corsini}, \&
  {Debattista}}]{Cuomo2020}
{Cuomo}, V., {Aguerri}, J.~A.~L., {Corsini}, E.~M., \& {Debattista}, V.~P.
  2020, \aap, 641, A111

\bibitem[{{Cuomo} {et~al.}(2019{\natexlab{a}}){Cuomo}, {Corsini}, {Aguerri},
  {Debattista}, {Coccato}, {Costantin}, {Dalla Bont{\`a}}, {Iodice},
  {M{\'e}ndez-Abreu}, {Morelli}, {Pagotto}, \& {Pizzella}}]{Cuomo2019b}
{Cuomo}, V., {Corsini}, E.~M., {Aguerri}, J.~A.~L., {et~al.}
  2019{\natexlab{a}}, \mnras, 488, 4972

\bibitem[{{Cuomo} {et~al.}(2022){Cuomo}, {Corsini}, {Morelli}, {Aguerri},
  {Lee}, {Coccato}, {Pizzella}, {Buttitta}, \& {Gasparri}}]{Cuomo2022}
{Cuomo}, V., {Corsini}, E.~M., {Morelli}, L., {et~al.} 2022, \mnras, 516, L24

\bibitem[{{Cuomo} {et~al.}(2021){Cuomo}, {Lee}, {Buttitta}, {Aguerri},
  {Corsini}, \& {Morelli}}]{Cuomo2021}
{Cuomo}, V., {Lee}, Y.~H., {Buttitta}, C., {et~al.} 2021, \aap, 649, A30

\bibitem[{{Cuomo} {et~al.}(2019{\natexlab{b}}){Cuomo}, {Lopez Aguerri},
  {Corsini}, {Debattista}, {M{\'e}ndez-Abreu}, \& {Pizzella}}]{Cuomo2019a}
{Cuomo}, V., {Lopez Aguerri}, J.~A., {Corsini}, E.~M., {et~al.}
  2019{\natexlab{b}}, \aap, 632, A51

\bibitem[{{Das} {et~al.}(2008){Das}, {Laurikainen}, {Salo}, \&
  {Buta}}]{Laurikainen2008}
{Das}, M., {Laurikainen}, E., {Salo}, H., \& {Buta}, R. 2008, \apss, 317, 163

\bibitem[{{Debattista}(2003)}]{deb2003}
{Debattista}, V.~P. 2003, \mnras, 342, 1194

\bibitem[{{Debattista} {et~al.}(2002){Debattista}, {Corsini}, \&
  {Aguerri}}]{Deb2002}
{Debattista}, V.~P., {Corsini}, E.~M., \& {Aguerri}, J.~A.~L. 2002, \mnras,
  332, 65

\bibitem[{{Debattista} \& {Sellwood}(2000)}]{debattista2000}
{Debattista}, V.~P. \& {Sellwood}, J.~A. 2000, \apj, 543, 704

\bibitem[{{Debattista} \& {Williams}(2004)}]{Deb2004}
{Debattista}, V.~P. \& {Williams}, T.~B. 2004, \apj, 605, 714

\bibitem[{{Dehnen} {et~al.}(2023){Dehnen}, {Semczuk}, \&
  {Sch{\"o}nrich}}]{newps}
{Dehnen}, W., {Semczuk}, M., \& {Sch{\"o}nrich}, R. 2023, \mnras, 518, 2712

\bibitem[{{England} {et~al.}(1990){England}, {Gottesman}, \&
  {Hunter}}]{england1990}
{England}, M.~N., {Gottesman}, S.~T., \& {Hunter}, Jr., J.~H. 1990, \apj, 348,
  456

\bibitem[{{Espejo Salcedo} {et~al.}(2025){Espejo Salcedo}, {Pastras},
  {V\textbackslash'acha}, {Pulsoni}, {Genzel}, {F\textbackslash''orster
  Schreiber}, {Jolly}, {Barfety}, {Chen}, {Tozzi}, {Liu}, {Lee}, {Wuyts},
  {Tacconi}, {Davies}, {\textbackslash''Ubler}, {Lutz}, {Wisnioski},
  {Shangguan}, {Lee}, {Sedona Price}, {Eisenhauer}, {Renzini}, {Nestor
  Shachar}, \& {Herrera-Camus}}]{2025arXiv250321738E}
{Espejo Salcedo}, J.~M., {Pastras}, S., {V\textbackslash'acha}, J., {et~al.}
  2025, arXiv e-prints, arXiv:2503.21738

\bibitem[{{Font} {et~al.}(2011){Font}, {Beckman}, {Epinat}, {Fathi},
  {Guti{\'e}rrez}, \& {Hernandez}}]{Font2011}
{Font}, J., {Beckman}, J.~E., {Epinat}, B., {et~al.} 2011, \apjl, 741, L14

\bibitem[{{Font} {et~al.}(2017){Font}, {Beckman}, {Mart{\'\i}nez-Valpuesta},
  {Borlaff}, {James}, {D{\'\i}az-Garc{\'\i}a}, {Garc{\'\i}a-Lorenzo},
  {Camps-Fari{\~n}a}, {Guti{\'e}rrez}, \& {Amram}}]{Font2017}
{Font}, J., {Beckman}, J.~E., {Mart{\'\i}nez-Valpuesta}, I., {et~al.} 2017,
  \apj, 835, 279

\bibitem[{{Font} {et~al.}(2014){Font}, {Beckman}, {Zaragoza-Cardiel}, {Fathi},
  {Epinat}, \& {Amram}}]{Font2014}
{Font}, J., {Beckman}, J.~E., {Zaragoza-Cardiel}, J., {et~al.} 2014, \mnras,
  444, L85

\bibitem[{{Fragkoudi} {et~al.}(2021){Fragkoudi}, {Grand}, {Pakmor}, {Springel},
  {White}, {Marinacci}, {Gomez}, \& {Navarro}}]{2021A&A...650L..16F}
{Fragkoudi}, F., {Grand}, R.~J.~J., {Pakmor}, R., {et~al.} 2021, \aap, 650, L16

\bibitem[{{Frankel} {et~al.}(2022){Frankel}, {Pillepich}, {Rix},
  {Rodriguez-Gomez}, {Sanders}, {Bovy}, {Kollmeier}, {Murray}, \&
  {Mackereth}}]{Frankel2022}
{Frankel}, N., {Pillepich}, A., {Rix}, H.-W., {et~al.} 2022, \apj, 940, 61

\bibitem[{{Frosst} {et~al.}(2025){Frosst}, {Obreschkow}, {Ludlow}, {Bottrell},
  \& {Genel}}]{math2025}
{Frosst}, M., {Obreschkow}, D., {Ludlow}, A., {Bottrell}, C., \& {Genel}, S.
  2025, \mnras

\bibitem[{{Gadotti}(2009)}]{2009ASSP....8..159G}
{Gadotti}, D.~A. 2009, in Astrophysics and Space Science Proceedings, Vol.~8,
  Chaos in Astronomy, ed. G.~{Contopoulos} \& P.~A. {Patsis}, 159

\bibitem[{{Garma-Oehmichen} {et~al.}(2020){Garma-Oehmichen}, {Cano-D{\'\i}az},
  {Hern{\'a}ndez-Toledo}, {Aquino-Ort{\'\i}z}, {Valenzuela}, {Aguerri},
  {S{\'a}nchez}, \& {Merrifield}}]{Garma2020}
{Garma-Oehmichen}, L., {Cano-D{\'\i}az}, M., {Hern{\'a}ndez-Toledo}, H.,
  {et~al.} 2020, \mnras, 491, 3655

\bibitem[{{Garma-Oehmichen} {et~al.}(2022){Garma-Oehmichen},
  {Hern{\'a}ndez-Toledo}, {Aquino-Ort{\'\i}z}, {Martinez-Medina}, {Puerari},
  {Cano-D{\'\i}az}, {Valenzuela}, {V{\'a}zquez-Mata}, {G{\'e}ron},
  {Mart{\'\i}nez-V{\'a}zquez}, \& {Lane}}]{Garma2022}
{Garma-Oehmichen}, L., {Hern{\'a}ndez-Toledo}, H., {Aquino-Ort{\'\i}z}, E.,
  {et~al.} 2022, \mnras, 517, 5660

\bibitem[{{Genel} {et~al.}(2015){Genel}, {Fall}, {Hernquist}, {Vogelsberger},
  {Snyder}, {Rodriguez-Gomez}, {Sijacki}, \& {Springel}}]{genel2015}
{Genel}, S., {Fall}, S.~M., {Hernquist}, L., {et~al.} 2015, \apjl, 804, L40

\bibitem[{{G{\'e}ron} {et~al.}(2025){G{\'e}ron}, {Smethurst}, {Dickinson},
  {Fortson}, {Garland}, {Kruk}, {Lintott}, {Makechemu}, {Mantha}, {Masters},
  {O'Ryan}, {Roberts}, {Simmons}, {Walmsley}, {Calabr{\`o}}, {Chiba},
  {Costantin}, {Drout}, {Fragkoudi}, {Guo}, {Holwerda}, {Jogee}, {Koekemoer},
  {Lucas}, \& {Pacucci}}]{luc2}
{G{\'e}ron}, T., {Smethurst}, R.~J., {Dickinson}, H., {et~al.} 2025, \apj, 987,
  74

\bibitem[{{G{\'e}ron} {et~al.}(2023){G{\'e}ron}, {Smethurst}, {Lintott},
  {Kruk}, {Masters}, {Simmons}, {Mantha}, {Walmsley}, {Garma-Oehmichen},
  {Drory}, \& {Lane}}]{Geron2023}
{G{\'e}ron}, T., {Smethurst}, R.~J., {Lintott}, C., {et~al.} 2023, \mnras, 521,
  1775

\bibitem[{{Gerssen} \& {Debattista}(2007)}]{Gerssen2007}
{Gerssen}, J. \& {Debattista}, V.~P. 2007, \mnras, 378, 189

\bibitem[{{Guo} {et~al.}(2019){Guo}, {Mao}, {Athanassoula}, {Li}, {Ge}, {Long},
  {Merrifield}, \& {Masters}}]{Guo2019}
{Guo}, R., {Mao}, S., {Athanassoula}, E., {et~al.} 2019, \mnras, 482, 1733

\bibitem[{{Guo} {et~al.}(2023){Guo}, {Jogee}, {Finkelstein}, {Chen}, {Wise},
  {Bagley}, {Barro}, {Wuyts}, {Kocevski}, {Kartaltepe}, {McGrath}, {Ferguson},
  {Mobasher}, {Giavalisco}, {Lucas}, {Zavala}, {Lotz}, {Grogin},
  {Huertas-Company}, {Vega-Ferrero}, {Hathi}, {Arrabal Haro}, {Dickinson},
  {Koekemoer}, {Papovich}, {Pirzkal}, {Yung}, {Backhaus}, {Bell},
  {Calabr{\`o}}, {Cleri}, {Coogan}, {Cooper}, {Costantin}, {Croton}, {Davis},
  {Dekel}, {Franco}, {Gardner}, {Holwerda}, {Hutchison}, {Pandya},
  {P{\'e}rez-Gonz{\'a}lez}, {Ravindranath}, {Rose}, {Trump}, {de la Vega}, \&
  {Wang}}]{mobasher2023}
{Guo}, Y., {Jogee}, S., {Finkelstein}, S.~L., {et~al.} 2023, \apjl, 945, L10

\bibitem[{{Guo} {et~al.}(2025){Guo}, {Jogee}, {Wise}, {Pritchett}, {McGrath},
  {Finkelstein}, {Iyer}, {Arrabal Haro}, {Bagley}, {Dickinson}, {Kartaltepe},
  {Koekemoer}, {Papovich}, {Pirzkal}, {Yung}, {Backhaus}, {Bell},
  {Bhatawdekar}, {Cheng}, {Costantin}, {de la Vega}, {Giavalisco}, {Hathi},
  {Holwerda}, {Kurczynski}, {Lucas}, {Mobasher}, {P{\'e}rez-Gonz{\'a}lez}, \&
  {Pacucci}}]{luc1}
{Guo}, Y., {Jogee}, S., {Wise}, E., {et~al.} 2025, \apj, 985, 181

\bibitem[{{Habibi} {et~al.}(2024){Habibi}, {Roshan}, {Hosseinirad},
  {Khosroshahi}, {Aguerri}, {Cuomo}, \& {Abbassi}}]{AHabibi}
{Habibi}, A., {Roshan}, M., {Hosseinirad}, M., {et~al.} 2024, \aap, 691, A122

\bibitem[{{Harborne} {et~al.}(2023){Harborne}, {Serene}, {Davies}, {Derkenne},
  {Vaughan}, {Burdon}, {Lagos}, {McDermid}, {O'Toole}, {Power}, {Robotham},
  {Santucci}, \& {Tobar}}]{harborne2023}
{Harborne}, K.~E., {Serene}, A., {Davies}, E.~J.~A., {et~al.} 2023, \pasa, 40,
  e048

\bibitem[{{Huertas-Company} {et~al.}(2025){Huertas-Company}, {Shuntov}, {Dong},
  {Walmsley}, {Ilbert}, {McCracken}, {Akins}, {Allen}, {Casey}, {Costantin},
  {Daddi}, {Dekel}, {Franco}, {Garland}, {G{\'e}ron}, {Gozaliasl},
  {Hirschmann}, {Kartaltepe}, {Koekemoer}, {Lintott}, {Liu}, {Lucas},
  {Masters}, {Pacucci}, {Paquereau}, {P'erez-Gonz'alez}, {Rhodes}, {Robertson},
  {Simmons}, {Smethurst}, {Toft}, \& {Yang}}]{2025arXiv250203532H}
{Huertas-Company}, M., {Shuntov}, M., {Dong}, Y., {et~al.} 2025, arXiv
  e-prints, arXiv:2502.03532

\bibitem[{{Ikhsanova} {et~al.}(2025){Ikhsanova}, {Costantin}, {Pizzella},
  {Corsini}, {Morelli}, {Ditrani}, {Ferr{\'e}-Mateu}, {Gabarra}, {Gullieuszik},
  {Haines}, {Iovino}, {Longhetti}, {Mercurio}, {Ragusa},
  {S{\'a}nchez-Bl{\'a}zquez}, {Tortora}, {Vulcani}, {Zhou}, {Gafton}, \&
  {Pistis}}]{2025arXiv250618997I}
{Ikhsanova}, A., {Costantin}, L., {Pizzella}, A., {et~al.} 2025, arXiv
  e-prints, arXiv:2506.18997

\bibitem[{{Jeong} {et~al.}(2007){Jeong}, {Bureau}, {Yi}, {Krajnovi{\'c}}, \&
  {Davies}}]{jeong2007}
{Jeong}, H., {Bureau}, M., {Yi}, S.~K., {Krajnovi{\'c}}, D., \& {Davies}, R.~L.
  2007, \mnras, 376, 1021

\bibitem[{{Jim{\'e}nez-Arranz} {et~al.}(2024){Jim{\'e}nez-Arranz}, {Chemin},
  {Romero-G{\'o}mez}, {Luri}, {Adamczyk}, {Castro-Ginard}, {Roca-F{\`a}brega},
  {McMillan}, \& {Cioni}}]{Arranz2024}
{Jim{\'e}nez-Arranz}, {\'O}., {Chemin}, L., {Romero-G{\'o}mez}, M., {et~al.}
  2024, \aap, 683, A102

\bibitem[{{Kormendy}(1979)}]{kormendy1979}
{Kormendy}, J. 1979, \apj, 227, 714

\bibitem[{{Krajnovi{\'c}} {et~al.}(2006){Krajnovi{\'c}}, {Cappellari}, {de
  Zeeuw}, \& {Copin}}]{pafit2}
{Krajnovi{\'c}}, D., {Cappellari}, M., {de Zeeuw}, P.~T., \& {Copin}, Y. 2006,
  \mnras, 366, 787

\bibitem[{{Krishnarao} {et~al.}(2022){Krishnarao}, {Pace}, {D'Onghia},
  {Aguerri}, {McClure}, {Peterken}, {Fern{\'a}ndez-Trincado}, {Merrifield},
  {Masters}, {Garma-Oehmichen}, {Boardman}, {Bershady}, {Drory}, \&
  {Lane}}]{Krishnarao2022}
{Krishnarao}, D., {Pace}, Z.~J., {D'Onghia}, E., {et~al.} 2022, \apj, 929, 112

\bibitem[{{Le Conte} {et~al.}(2024){Le Conte}, {Gadotti}, {Ferreira},
  {Conselice}, {de S{\'a}-Freitas}, {Kim}, {Neumann}, {Fragkoudi},
  {Athanassoula}, \& {Adams}}]{bar2024_2}
{Le Conte}, Z.~A., {Gadotti}, D.~A., {Ferreira}, L., {et~al.} 2024, \mnras,
  530, 1984

\bibitem[{{Liang} {et~al.}(2024){Liang}, {Yu}, {Fang}, \& {Ho}}]{liang}
{Liang}, X., {Yu}, S.-Y., {Fang}, T., \& {Ho}, L.~C. 2024, \aap, 688, A158

\bibitem[{{Lin} {et~al.}(2013){Lin}, {Wang}, {Hsieh}, {Taam}, {Yang}, \&
  {Yen}}]{lin2013}
{Lin}, L.-H., {Wang}, H.-H., {Hsieh}, P.-Y., {et~al.} 2013, \apj, 771, 8

\bibitem[{{Lindblad} \& {Kristen}(1996)}]{lindblad1996}
{Lindblad}, P.~A.~B. \& {Kristen}, H. 1996, \aap, 313, 733

\bibitem[{{Marinacci} {et~al.}(2018){Marinacci}, {Vogelsberger}, {Pakmor},
  {Torrey}, {Springel}, {Hernquist}, {Nelson}, {Weinberger}, {Pillepich},
  {Naiman}, \& {Genel}}]{tng5}
{Marinacci}, F., {Vogelsberger}, M., {Pakmor}, R., {et~al.} 2018, \mnras, 480,
  5113

\bibitem[{{M{\'e}ndez-Abreu} {et~al.}(2023){M{\'e}ndez-Abreu}, {Costantin}, \&
  {Kruk}}]{2023A&A...678A..54M}
{M{\'e}ndez-Abreu}, J., {Costantin}, L., \& {Kruk}, S. 2023, \aap, 678, A54

\bibitem[{{Mu{\~n}oz-Tu{\~n}{\'o}n} {et~al.}(2004){Mu{\~n}oz-Tu{\~n}{\'o}n},
  {Caon}, \& {Aguerri}}]{Mu2004}
{Mu{\~n}oz-Tu{\~n}{\'o}n}, C., {Caon}, N., \& {Aguerri}, J. A.~L. 2004, \aj,
  127, 58

\bibitem[{{Naiman} {et~al.}(2018){Naiman}, {Pillepich}, {Springel},
  {Ramirez-Ruiz}, {Torrey}, {Vogelsberger}, {Pakmor}, {Nelson}, {Marinacci},
  {Hernquist}, {Weinberger}, \& {Genel}}]{tng4}
{Naiman}, J.~P., {Pillepich}, A., {Springel}, V., {et~al.} 2018, \mnras, 477,
  1206

\bibitem[{{Nanni} {et~al.}(2022){Nanni}, {Thomas}, {Trayford}, {Maraston},
  {Neumann}, {Law}, {Hill}, {Pillepich}, {Yan}, {Chen}, \&
  {Lazarz}}]{nanni2022}
{Nanni}, L., {Thomas}, D., {Trayford}, J., {et~al.} 2022, \mnras, 515, 320

\bibitem[{{Nelson} {et~al.}(2019{\natexlab{a}}){Nelson}, {Pillepich},
  {Springel}, {Pakmor}, {Weinberger}, {Genel}, {Torrey}, {Vogelsberger},
  {Marinacci}, \& {Hernquist}}]{TNG502}
{Nelson}, D., {Pillepich}, A., {Springel}, V., {et~al.} 2019{\natexlab{a}},
  \mnras, 490, 3234

\bibitem[{{Nelson} {et~al.}(2018){Nelson}, {Pillepich}, {Springel},
  {Weinberger}, {Hernquist}, {Pakmor}, {Genel}, {Torrey}, {Vogelsberger},
  {Kauffmann}, {Marinacci}, \& {Naiman}}]{tng3}
{Nelson}, D., {Pillepich}, A., {Springel}, V., {et~al.} 2018, \mnras, 475, 624

\bibitem[{{Nelson} {et~al.}(2019{\natexlab{b}}){Nelson}, {Springel},
  {Pillepich}, {Rodriguez-Gomez}, {Torrey}, {Genel}, {Vogelsberger}, {Pakmor},
  {Marinacci}, {Weinberger}, {Kelley}, {Lovell}, {Diemer}, \&
  {Hernquist}}]{TNG501}
{Nelson}, D., {Springel}, V., {Pillepich}, A., {et~al.} 2019{\natexlab{b}},
  Computational Astrophysics and Cosmology, 6, 2

\bibitem[{{Ohta} {et~al.}(1990){Ohta}, {Hamabe}, \& {Wakamatsu}}]{ohta}
{Ohta}, K., {Hamabe}, M., \& {Wakamatsu}, K.-I. 1990, \apj, 357, 71

\bibitem[{{P{\'e}rez} {et~al.}(2012){P{\'e}rez}, {Aguerri}, \&
  {M{\'e}ndez-Abreu}}]{2012A&A...540A.103P}
{P{\'e}rez}, I., {Aguerri}, J.~A.~L., \& {M{\'e}ndez-Abreu}, J. 2012, \aap,
  540, A103

\bibitem[{{Perrin} {et~al.}(2014){Perrin}, {Sivaramakrishnan}, {Lajoie},
  {Elliott}, {Pueyo}, {Ravindranath}, \& {Albert}}]{WebbPSF}
{Perrin}, M.~D., {Sivaramakrishnan}, A., {Lajoie}, C.-P., {et~al.} 2014, in
  Society of Photo-Optical Instrumentation Engineers (SPIE) Conference Series,
  Vol. 9143, Space Telescopes and Instrumentation 2014: Optical, Infrared, and
  Millimeter Wave, ed. J.~M. {Oschmann}, Jr., M.~{Clampin}, G.~G. {Fazio}, \&
  H.~A. {MacEwen}, 91433X

\bibitem[{{Pillepich} {et~al.}(2018){Pillepich}, {Nelson}, {Hernquist},
  {Springel}, {Pakmor}, {Torrey}, {Weinberger}, {Genel}, {Naiman}, {Marinacci},
  \& {Vogelsberger}}]{tng1}
{Pillepich}, A., {Nelson}, D., {Hernquist}, L., {et~al.} 2018, \mnras, 475, 648

\bibitem[{{Pillepich} {et~al.}(2019){Pillepich}, {Nelson}, {Springel},
  {Pakmor}, {Torrey}, {Weinberger}, {Vogelsberger}, {Marinacci}, {Genel}, {van
  der Wel}, \& {Hernquist}}]{TNG503}
{Pillepich}, A., {Nelson}, D., {Springel}, V., {et~al.} 2019, \mnras, 490, 3196

\bibitem[{{Rautiainen} {et~al.}(2008){Rautiainen}, {Salo}, \&
  {Laurikainen}}]{Rautiainen2008}
{Rautiainen}, P., {Salo}, H., \& {Laurikainen}, E. 2008, \mnras, 388, 1803

\bibitem[{{Rimoldini}(2014)}]{rimoldini2014}
{Rimoldini}, L. 2014, Astronomy and Computing, 5, 1

\bibitem[{{Rosas-Guevara} {et~al.}(2022){Rosas-Guevara}, {Bonoli}, {Dotti},
  {Izquierdo-Villalba}, {Lupi}, {Zana}, {Bonetti}, {Nelson}, {Springel},
  {Hernquist}, \& {Vogelsberger}}]{2022MNRAS.512.5339R}
{Rosas-Guevara}, Y., {Bonoli}, S., {Dotti}, M., {et~al.} 2022, \mnras, 512,
  5339

\bibitem[{{Roshan} {et~al.}(2021){Roshan}, {Ghafourian}, {Kashfi}, {Banik},
  {Haslbauer}, {Cuomo}, {Famaey}, \& {Kroupa}}]{2021MNRAS.508..926R}
{Roshan}, M., {Ghafourian}, N., {Kashfi}, T., {et~al.} 2021, \mnras, 508, 926

\bibitem[{{Sales} {et~al.}(2010){Sales}, {Navarro}, {Schaye}, {Dalla Vecchia},
  {Springel}, \& {Booth}}]{sales2010}
{Sales}, L.~V., {Navarro}, J.~F., {Schaye}, J., {et~al.} 2010, \mnras, 409,
  1541

\bibitem[{{Schaye} {et~al.}(2015){Schaye}, {Crain}, {Bower}, {Furlong},
  {Schaller}, {Theuns}, {Dalla Vecchia}, {Frenk}, {McCarthy}, {Helly},
  {Jenkins}, {Rosas-Guevara}, {White}, {Baes}, {Booth}, {Camps}, {Navarro},
  {Qu}, {Rahmati}, {Sawala}, {Thomas}, \& {Trayford}}]{eagle}
{Schaye}, J., {Crain}, R.~A., {Bower}, R.~G., {et~al.} 2015, \mnras, 446, 521

\bibitem[{{Sellwood}(2014{\natexlab{a}})}]{2014arXiv1406.6606S}
{Sellwood}, J.~A. 2014{\natexlab{a}}, arXiv e-prints, arXiv:1406.6606

\bibitem[{{Sellwood}(2014{\natexlab{b}})}]{2014RvMP...86....1S}
{Sellwood}, J.~A. 2014{\natexlab{b}}, Reviews of Modern Physics, 86, 1

\bibitem[{{Semczuk} {et~al.}(2024){Semczuk}, {Dehnen}, {Sch{\"o}nrich}, \&
  {Athanassoula}}]{dehnen2024}
{Semczuk}, M., {Dehnen}, W., {Sch{\"o}nrich}, R., \& {Athanassoula}, E. 2024,
  \aap, 692, A159

\bibitem[{{Sempere} {et~al.}(1995){Sempere}, {Garcia-Burillo}, {Combes}, \&
  {Knapen}}]{Sempere1995}
{Sempere}, M.~J., {Garcia-Burillo}, S., {Combes}, F., \& {Knapen}, J.~H. 1995,
  \aap, 296, 45

\bibitem[{{Sheth} {et~al.}(2012){Sheth}, {Melbourne}, {Elmegreen}, {Elmegreen},
  {Athanassoula}, {Abraham}, \& {Weiner}}]{2012ApJ...758..136S}
{Sheth}, K., {Melbourne}, J., {Elmegreen}, D.~M., {et~al.} 2012, \apj, 758, 136

\bibitem[{{Smail} {et~al.}(2023){Smail}, {Dudzevi{\v{c}}i{\={u}}t{\.{e}}},
  {Gurwell}, {Fazio}, {Willner}, {Swinbank}, {Arumugam}, {Summers}, {Cohen},
  {Jansen}, {Windhorst}, {Meena}, {Zitrin}, {Keel}, {Cheng}, {Coe},
  {Conselice}, {D'Silva}, {Driver}, {Frye}, {Grogin}, {Koekemoer}, {Marshall},
  {Nonino}, {Pirzkal}, {Robotham}, {Rutkowski}, {Ryan}, {Tompkins}, {Willmer},
  {Yan}, {Broadhurst}, {Diego}, {Kamieneski}, \& {Yun}}]{bar_2023}
{Smail}, I., {Dudzevi{\v{c}}i{\={u}}t{\.{e}}}, U., {Gurwell}, M., {et~al.}
  2023, \apj, 958, 36

\bibitem[{{Springel} {et~al.}(2018){Springel}, {Pakmor}, {Pillepich},
  {Weinberger}, {Nelson}, {Hernquist}, {Vogelsberger}, {Genel}, {Torrey},
  {Marinacci}, \& {Naiman}}]{tng2}
{Springel}, V., {Pakmor}, R., {Pillepich}, A., {et~al.} 2018, \mnras, 475, 676

\bibitem[{{Tremaine} \& {Weinberg}(1984)}]{Tremaine1984}
{Tremaine}, S. \& {Weinberg}, M.~D. 1984, \apjl, 282, L5

\bibitem[{{Tsukui} {et~al.}(2024){Tsukui}, {Wisnioski}, {Bland-Hawthorn},
  {Mai}, {Iguchi}, {Baba}, \& {Freeman}}]{bar2024_1}
{Tsukui}, T., {Wisnioski}, E., {Bland-Hawthorn}, J., {et~al.} 2024, \mnras,
  527, 8941

\bibitem[{{Weiner} {et~al.}(2001){Weiner}, {Sellwood}, \&
  {Williams}}]{weiner2001}
{Weiner}, B.~J., {Sellwood}, J.~A., \& {Williams}, T.~B. 2001, \apj, 546, 931

\bibitem[{{Zana} {et~al.}(2022){Zana}, {Lupi}, {Bonetti}, {Dotti},
  {Rosas-Guevara}, {Izquierdo-Villalba}, {Bonoli}, {Hernquist}, \&
  {Nelson}}]{2022MNRAS.515.1524Z}
{Zana}, T., {Lupi}, A., {Bonetti}, M., {et~al.} 2022, \mnras, 515, 1524

\bibitem[{{Zhang} {et~al.}(2024){Zhang}, {Belokurov}, {Evans}, {Kane}, \&
  {Sanders}}]{zhang2024}
{Zhang}, H., {Belokurov}, V., {Evans}, N.~W., {Kane}, S.~G., \& {Sanders},
  J.~L. 2024, \mnras, 533, 3395

\bibitem[{{Zhang}(1998)}]{1998ApJ...499...93Z}
{Zhang}, X. 1998, \apj, 499, 93

\bibitem[{{Zhang} \& {Buta}(2007)}]{zhang2007}
{Zhang}, X. \& {Buta}, R.~J. 2007, \aj, 133, 2584

\bibitem[{{Zou} {et~al.}(2019){Zou}, {Shen}, {Bureau}, \& {Li}}]{Zou2019}
{Zou}, Y., {Shen}, J., {Bureau}, M., \& {Li}, Z.-Y. 2019, \apj, 884, 23

\end{thebibliography}

\begin{appendix}
\section{A quick test for DSS method using the {\small GALAXY} code}
\label{test}
To evaluate the accuracy of the DSS method, we conduct an N-body simulation using the {\small GALAXY} code \citep{2014arXiv1406.6606S}. In this simulation, we implement a baryonic exponential disk
\begin{equation}\label{disk_den}
	\rho(R,z)=\frac{M_\mathrm{d}}{4\pi z_\mathrm{d} R_\mathrm{d}^2} e^{-\frac{R}{R_\mathrm{d}}} \text{sech}^2\left(\frac{z}{2z_\mathrm{d}}\right)
\end{equation}
 characterized by a length scale of $R_\mathrm{d} = 1.13 \, \text{kpc}$, the vertical scale height of $z_\mathrm{d} = 0.09 \, \text{kpc}$, and the mass of $M_\mathrm{d} = 1.8 \times 10^{10} \, M_{\odot}$. In addition, the spherical dark matter halo is modeled using a Plummer profile:
\begin{eqnarray}\label{halo_den}
	\rho_{\mathrm{h}}(r)=\frac{M_\mathrm{h}}{4\pi r_\mathrm{h}^3}\Big(1+\frac{r^2}{r_\mathrm{h}^2}\Big)^{-5/2}.
\end{eqnarray}
The halo has a mass of $M_\mathrm{h} = 2.83 M_\mathrm{d}$ and a length scale of $r_\mathrm{h} = 11.2 R_\mathrm{d}$. In this configuration, a bar structure rapidly forms. Once the system reaches a stable state, we measure the pattern speed using the DSS method. Notably, the {\small GALAXY} code provides precise measurements of the pattern speed over time. For the DSS method, we only require the face-on view of the galaxy and the positions and velocity of the particles.

\begin{figure}
  \centering
  \includegraphics[width=0.40\textwidth]{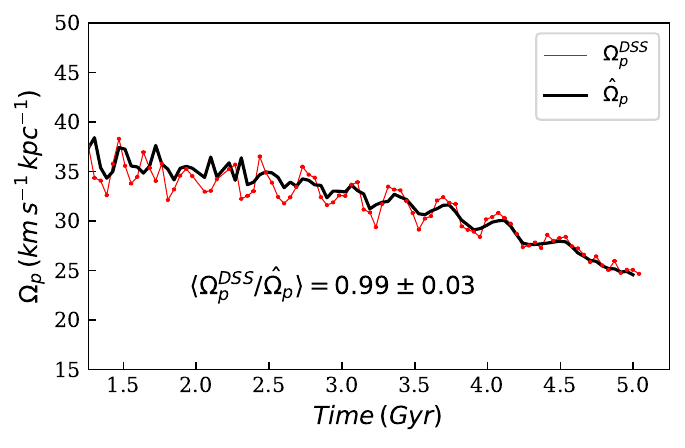}
  \caption{Time evolution of the pattern speed for the isolated simulation presented in Appendix \ref{test}. The intrinsic pattern speed, $\hat{\Omega}_{\rm p}$ (black line), is obtained from the {\small GALAXY} code; the pattern speed is calculated using the DSS method, $\Omega^{\text{DSS}}_{\rm p}$ (red line). We analyzed 95 snapshots from the simulation, and the figure also includes the mean value and corresponding standard deviation of the ratio $\Omega^{\text{DSS}}_{\rm p}/\hat{\Omega}_{\rm p}$, providing a quantitative comparison of the DSS method accuracy relative to the results from the {\small GALAXY} code.}
 \label{fig1}
\end{figure}

Figure \ref{fig1} illustrates the time evolution of the pattern speed. The black line represents the intrinsic $\hat{\Omega}_{\rm p}$ obtained from the {\small GALAXY} code. We observe that the bar is decelerating due to angular momentum exchange between the bar and dark matter halo. The red line depicts the pattern speed, $\Omega^{\text{DSS}}_{\rm p}$, calculated using the DSS method. We analyze 95 snapshots from the simulation. For each snapshot, we compute the ratio $\Omega^{\text{DSS}}_{\rm p}/\hat{\Omega}_{\rm p}$. We then calculate the mean value of this ratio across all snapshots, using the standard deviation as an estimate of the error. The result is
\begin{equation}
\langle\Omega^{\text{DSS}}_{\rm p}/\hat{\Omega}_{\rm p}\rangle=0.99\pm 0.03 .
\end{equation}
This result demonstrates that the DSS method performs in an excellent way. 

\section{Measuring bar pattern speeds with JWST: Redshift limits}
\label{JWST_TW}
\begin{table*}[h!]
\centering
\caption{NIRSpec capability for measuring bar pattern speeds using the TW method}
\label{tab:beautiful_table}
\begin{tabular}{ccccc}
\toprule
Filter & Wavelength range & FWHM$_\mathrm{PSF}$ & Redshift range & $l_\mathrm{b}^\mathrm{min}$ \\
       & ($\mu$m)          & ($^{\prime\prime}$) &                & (kpc)                       \\
\midrule
F070LP & 0.7 – 1.27        & 0.127               & $0.35 \lesssim z \lesssim 1.45$ & 2.90 \\
F100LP & 0.97 – 1.89       & 0.128               & $0.87 \lesssim z \lesssim 2.65$ & 3.26 \\
F170LP & 1.66 – 3.17       & 0.138               & $2.21 \lesssim z \lesssim 5.13$ & 3.01 \\
F290LP & 2.87 – 5.27       & 0.170               & $4.55 \lesssim z \lesssim 9.19$ & 2.77 \\
\bottomrule
\end{tabular}
\tablefoot{Redshift ranges where NIRSpec filters can detect barred galaxies suitable for the TW method. These intervals are derived under the assumptions that the Mg line triplet serves as the primary absorption line used in the TW method and inclination angle is $45^{\circ}$. The PSF size is used to determine the minimum required bar length, $l_\mathrm{b}^{\rm min}$, for detection by the TW method. Since both the PSF size and $l_\mathrm{b}^{\rm min}$ vary with redshift, we report their mean values for each filter. As mentioned in the text, we use simulated PSFs obtained from STPSF software.}
\label{tab1}
\end{table*}

\begin{figure}
\centering
\includegraphics[width=0.33\textwidth]{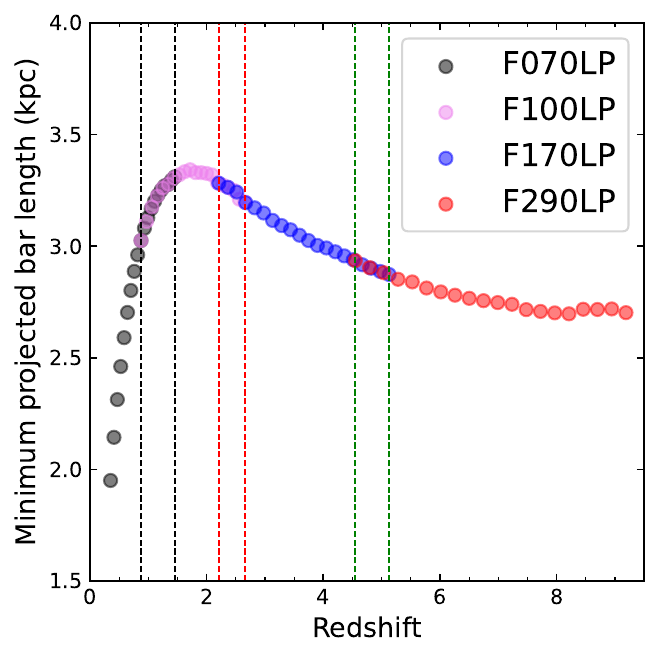}
\caption{Minimum projected bar length, $l_\mathrm{b}^{\rm min}$, required for the TW method as a function of redshift. The vertical lines indicate the redshift ranges where two successive filters overlap. As indicated in the legends, the different colors on the main curve, represent different filters.}
 \label{l_b_min}
\end{figure}
It is intriguing to derive the redshifts at which JWST can facilitate the application of the TW method to measure pattern speeds. Upon closer examination, it becomes clear that different NIRSpec filters cover distinct redshift ranges suitable for this method. To quantify this, we select the Mg line triplet, situated approximately within the wavelength range $0.5167-0.5183\, \mu$m, as it is one of the most significant stellar absorption feature commonly employed in the TW method. Since different filters of NIRSpec span different wavelength ranges, we determine the redshift range where the Mg line triplet falls within the wavelength range of each filter. The results are presented in the fifth column of Table \ref{tab1}.This is a theoretical estimation, where we disregard technical issues such as exposure time and $S/N$.

In the fourth column of Table \ref{tab1}, we report the ${\rm FWHM_{PSF}}$. Since the PSF size varies with wavelength, we consider its mean value for each filter. In the sixth column, we specify the minimum required value of the projected bar length $l_\mathrm{b}$, $l_\mathrm{b}^{\rm min}$, for detection using the TW method. More specifically, because the spatial sampling ($0\rlap{\raisebox{0.5ex}{$^{\prime\prime}$}}.1$\footnote{This value can be improved depending on the numbers and design of dithering patterns. In this work, we adopt a conservative estimate by assuming no dithering.}) is smaller than the PSF size, we have $l_\mathrm{b}^{\rm min}=3 \times {\rm FWHM_{PSF}}$. This minimum value varies with redshift, so we have recorded a mean value for each filter. In Fig. \ref{l_b_min}, we illustrate the $l_b^{\rm min}(z)$ as a function of redshift for different filters. The vertical lines indicate the redshift ranges where two successive filters overlap. The existence of a maximum in $l_\mathrm{b}^{\rm min}$ is directly attributed to the behavior of the angular diameter distance, $d_A(z)$, in standard cosmology. The function $d_A(z)$ reaches a maximum around $z \simeq 1.6$ and decreases as a function of $z$ beyond this redshift. In the above analysis, we estimate $d_A(z)$ employing the cosmological parameters listed in Section \ref{intro}.

Although our analysis extends to redshift beyond $z \sim 4$, it is important to note that, based on galaxy formation scenarios in standard cosmology, the existence of bars at those redshifts is unlikely. In other words, galactic disks may not be dynamically cold or massive enough to have formed a stellar bar \citep{2012ApJ...758..136S}. On the other hand, to date, all observed bars have redshifts $z<4$ \citep{mobasher2023,2023Natur.623..499C,2025MNRAS.536.3757A}. Therefore, if we conservatively limit ourselves to $z\lesssim 5$, the F170LP filter remains capable of detecting all ancient barred galaxies with projected bar lengths greater than $\sim 3$ kpc. These bars would be viable candidates to be measures by the TW method.

It is essential to note that, in this paper, our analysis is restricted to the medium resolution F070LP filter with bars at a redshift of $z\simeq 1.2$. Although the results may remain valid for the F100LP filter, where the PSF size is not significantly different, further investigation is necessary for the F170LP filter.It may be helpful to note that the analysis presented in this paper can be replicated for application with the F170LP filter.

\end{appendix}
\end{document}